\documentclass[twocolumn,english,aps,prb,twocolum,superscriptaddress,nobibnotes,amsmath,amssymb,floatfix,longbibliography]{revtex4-1}

\makeatletter
\renewcommand\@make@capt@title[2]{%
\@ifx@empty\float@link{\@firstofone}{\expandafter\href\expandafter{\float@link}}%
{\textbf{#1}}\@caption@fignum@sep#2\quad}%
\makeatother
 
\makeatletter 
\renewcommand{\fnum@figure}{\textbf{Figure~\thefigure}}
\makeatother

\usepackage[utf8]{inputenc}
\usepackage[T1]{fontenc}
\usepackage[english]{babel}
\usepackage{amsmath,amsfonts,amssymb}
\usepackage{lmodern}
\usepackage{upgreek}
\usepackage{url}
\usepackage{placeins}
\usepackage{ragged2e}
\usepackage{subcaption}
\DeclareCaptionLabelFormat{bold}{\textbf{(#2)}}
\DeclareCaptionJustification{justified}{\justifying}
\usepackage{hyperref}

\newcommand{\SIref}[1]{Supplementary Note~\ref{#1}}

\usepackage[version=3]{mhchem}
\usepackage[squaren, Gray, mediumqspace]{SIunits}

\usepackage{epstopdf}
\usepackage{graphicx}
\graphicspath{{./Figures/}}

\usepackage{titlesec}
\titleformat{\section}{\normalfont\large\bfseries}{\thesection}{1em}{}
\titlespacing{\section}{0pt}{10pt}{5pt minus 5pt}
\titleformat{\subsection}[runin]{\normalfont\bfseries}{\thesubsection}{0pt}{}[\mbox{ -- }]
\titlespacing{\subsection}{0pt}{8pt plus 5pt minus 5pt}{0pt}

\newcommand{\fref}[1]{Fig.~\ref{#1}}

\def \fceo {f_{\textsc{ceo}}}
\def \frep {f_{\rm rep}}
\def \fkrep {f_{\rm rep}^{\rm K}}
\def \fauxrep {f_{\rm rep}^{\rm aux}}
\def \fkceo {f_{\textsc{ceo}}^{\rm K}}
\def \fauxceo {f_{\textsc{ceo}}^{\rm aux}}

\begin{document}

\title{Ultralow-Noise Photonic Microwave Synthesis using a Soliton Microcomb-based Transfer Oscillator}

\author{Erwan~Lucas}
\thanks{These authors contributed equally to this work.}
\affiliation{Institute of Physics, École Polytechnique Fédérale de Lausanne (EPFL), CH-1015 Lausanne, Switzerland}

\author{Pierre~Brochard}
\thanks{These authors contributed equally to this work.}
\affiliation{Laboratoire Temps-Fréquence, Université de Neuchâtel, CH-2000 Neuchâtel, Switzerland}

\author{Romain~Bouchand}
\affiliation{Institute of Physics, École Polytechnique Fédérale de Lausanne (EPFL), CH-1015 Lausanne, Switzerland}

\author{St\'ephane~Schilt}
\affiliation{Laboratoire Temps-Fréquence, Université de Neuchâtel, CH-2000 Neuchâtel, Switzerland}

\author{Thomas~Südmeyer}
\affiliation{Laboratoire Temps-Fréquence, Université de Neuchâtel, CH-2000 Neuchâtel, Switzerland}

\author{Tobias~J.~Kippenberg}
\email[]{tobias.kippenberg@epfl.ch}
\affiliation{Institute of Physics, École Polytechnique Fédérale de Lausanne (EPFL), CH-1015 Lausanne, Switzerland}

\maketitle

\noindent\textbf{\boldmath
The synthesis of ultralow-noise microwaves is of both scientific and technological relevance for timing, metrology, communications and radio-astronomy. Today, the lowest reported phase noise signals are obtained via optical frequency-division using mode-locked laser frequency combs. Nonetheless, this technique ideally requires high repetition rates and tight comb stabilisation.
Here, a soliton microcomb with a 14~GHz repetition rate is generated with an ultra-stable pump laser and used to derive an ultralow-noise microwave reference signal, with an absolute phase noise level below \unit{-60}{dBc/Hz} at 1~Hz offset frequency and \unit{-135}{dBc/Hz} at 10~kHz.
This is achieved using a transfer oscillator approach, where the free-running microcomb noise (which is carefully studied and minimised) is cancelled via a combination of electronic division and mixing. Although this proof-of-principle uses an auxiliary comb for detecting the microcomb's offset frequency, we highlight the prospects of this method with future self-referenced integrated microcombs and electro-optic combs, that would allow for ultralow-noise microwave and sub-terahertz signal generators.
}

\begin{figure*}[t]
\centering
\includegraphics[width=\textwidth]{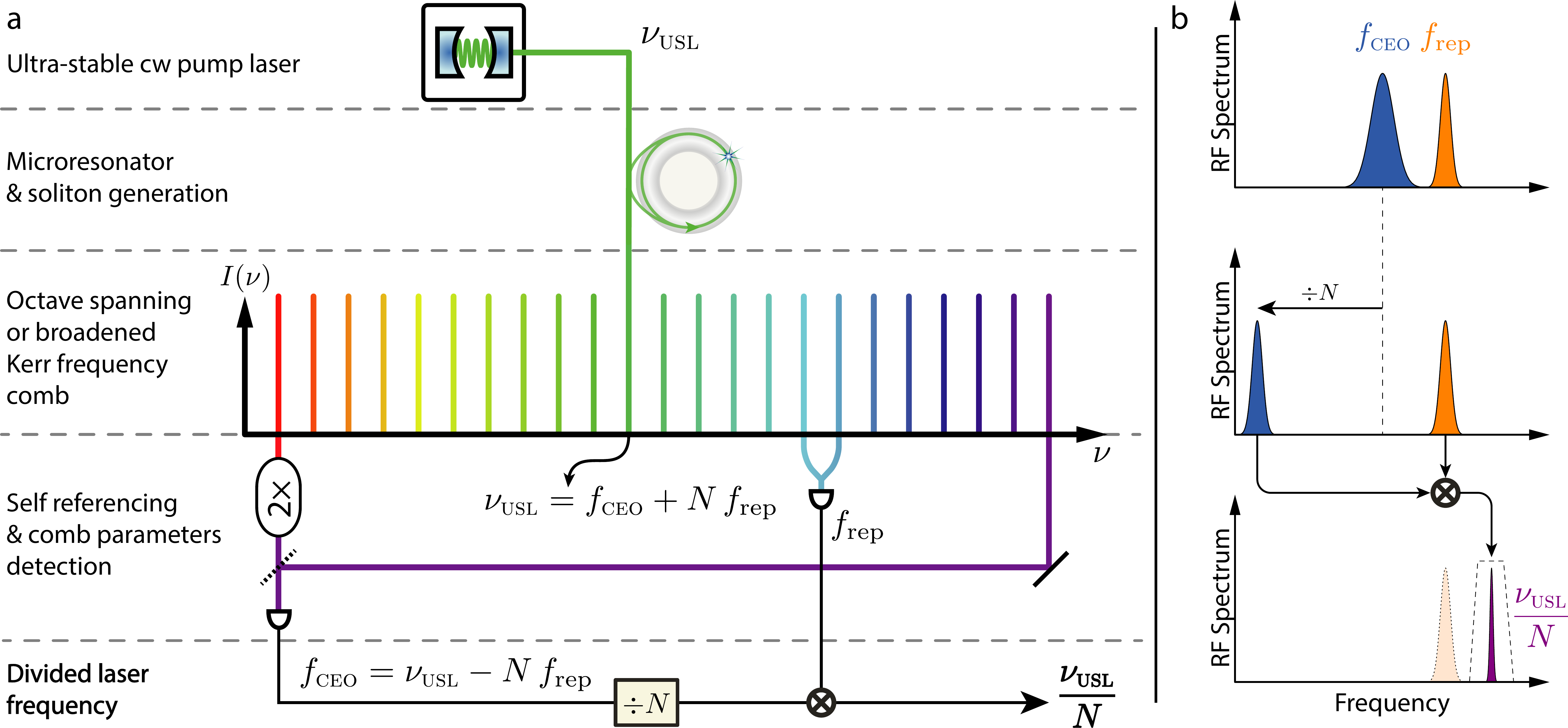}
{\phantomsubcaption\label{fig:concept:mockup}}
{\phantomsubcaption\label{fig:concept:spectrum}}
\caption{\textbf{Principle of operation of the Kerr comb-based transfer oscillator} for optical-to-microwave frequency division.
\subref{fig:concept:mockup} Schematic illustration of the transfer oscillator applied to a Kerr comb (or electro-optic combs equivalently).
\subref{fig:concept:spectrum} Schematic representation of the signal evolution along the electronic division chain leading to the low-noise output signal. The two comb parameters $\fceo$ and $\frep$ are detected. Both parameters can be free-running and fluctuate. The carrier envelope offset (CEO) frequency is electronically divided by a large number $N$ that corresponds to the tooth number of the ultra-stable pump $ \nu_{\textsc{usl}} $. After this step, the frequency fluctuations of the divided CEO $\fceo / N = \nu_{\textsc{usl}} / N - \frep$ are dominated by the repetition rate fluctuations. These are removed by mixing $\fceo / N$ with $\frep$ to obtain the division result $\nu_{\textsc{usl}} / N$. A narrow-band filtering is used to reject spurs.
}
\label{fig:concept}
\end{figure*}

\begin{figure*}[t]
\centering
\includegraphics[width=\textwidth]{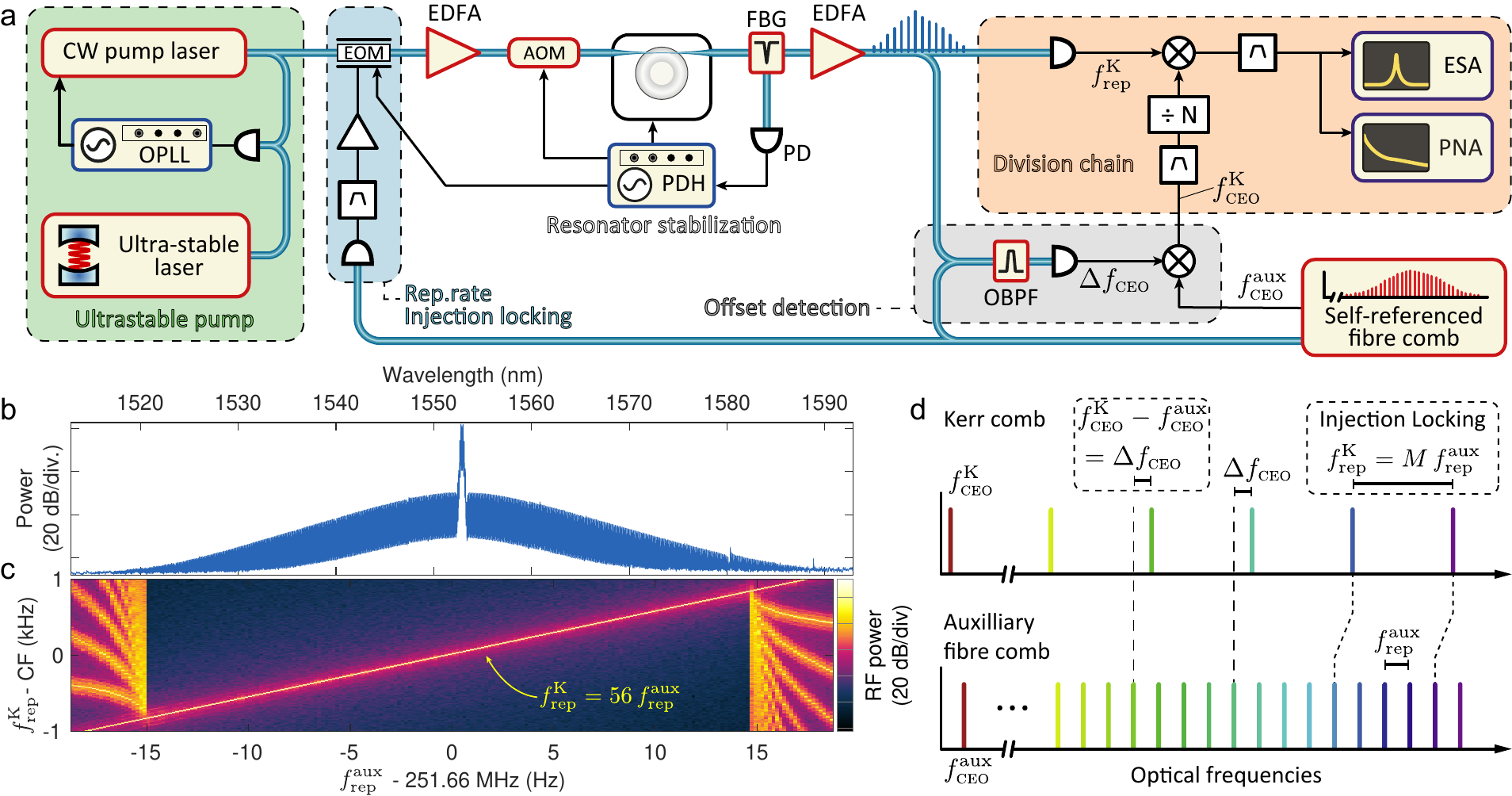}
{\phantomsubcaption\label{fig:setup:setup}}
{\phantomsubcaption\label{fig:setup:Optspectrum}}
{\phantomsubcaption\label{fig:setup:inj_lock_range}}
{\phantomsubcaption\label{fig:setup:injlock}}
\caption{\textbf{Experimental setup and CEO detection with the auxiliary comb}
\subref{fig:setup:setup} Setup for Kerr comb-based optical frequency division. The details of each highlighted block can be found in the supplementary information. EDFA, Er-doped fibre amplifier; AOM, Acousto-optic modulator; EOM, Electro-optic modulator; FBG, Fibre Bragg grating for pump rejection; OBPF, Optical band-pass filter; PD, Photodiode; OPLL, Optical phase lock loop; PDH, Pound-Drever-Hall lock; PNA, Phase noise analyser; ESA, Electrical spectrum analyser.
\subref{fig:setup:Optspectrum} Optical spectrum of the soliton-based Kerr comb, prior to pump suppression.
\subref{fig:setup:inj_lock_range}
Radio-frequency (RF) spectrogram showing the injection-locking effect of the Kerr comb repetition rate $\fkrep$ to the 56\textsuperscript{th} harmonic of the auxiliary comb repetition rate $\fauxrep$, obtained by changing the frequency of $\fauxrep$. Here, the harmonic power applied to the EOM is $\sim 11$~dBm yielding a locking range of $ \sim 1.7 $~kHz. The spectrum of $\fkrep$ is centred at $ \text{CF} = 14.092943$~GHz and the resolution bandwidth is 5~Hz.
\subref{fig:setup:injlock} Principle of the Kerr comb CEO detection with the auxiliary comb. The harmonic relation between the repetition rate of both combs is ensured via injection locking for $M=56$. The heterodyne beat between the two combs thus yields the difference between their carrier-envelope offset frequency ($\Delta f_{\textsc{ceo}}$).
}
\label{fig:setup}
\end{figure*}

\begin{figure*}
	\centering
	\includegraphics[width=\textwidth]{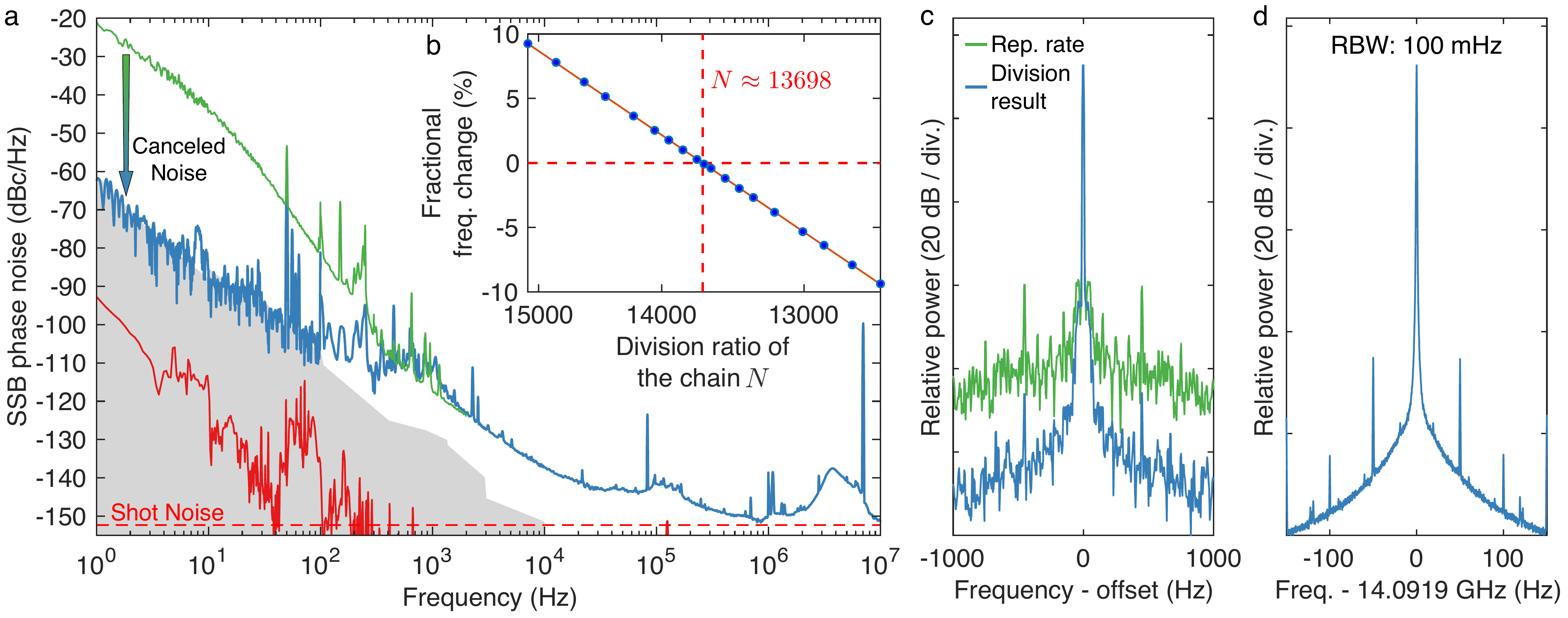}
	{\phantomsubcaption\label{fig:result:PN}}
	{\phantomsubcaption\label{fig:result:ratio}}
	{\phantomsubcaption\label{fig:result:RF_spectr_comp}}
	{\phantomsubcaption\label{fig:result:Fine_RF_Spectr}}
	\caption{\textbf{Characterisation of the optical-to-microwave division signal}
		\subref{fig:result:PN} Absolute single-sideband (SSB) phase noise of the 14.09~GHz signal generated by optical-to-microwave division of the USL via the Kerr comb transfer oscillator (blue) and obtained directly from the Kerr comb repetition rate (green) for comparison. The sensitivity limit of the phase noise analyser (3000 cross correlations applied at 1~Hz) is indicated by the grey shaded area. The red line is the limit inferred from the optical phase noise of the USL, assuming an ideal noiseless division.
		\subref{fig:result:ratio} Precise determination of the optimal division factor $N$ corresponding to the zero crossing of the linear fit (solid line) of the measured relative frequency change of the generated RF signal for a small variation of the repetition rate (dots).
		\subref{fig:result:RF_spectr_comp} Comparison between the RF spectra of the Kerr comb repetition rate and the optical-to-microwave frequency division result. The resolution bandwidth (RBW) is 5~Hz.
		\subref{fig:result:Fine_RF_Spectr} RF spectrum of the frequency-divided output signal, the RBW is 100~mHz. The data was acquired with the IQ demodulation mode of the spectrum analyser.
	}
	\label{fig:result}
\end{figure*}

\section*{Introduction}

The synthesis of microwave signals via photonic systems, such as dual frequency lasers~\cite{Pillet2008}, optoelectronic oscillators~\cite{Maleki2011}, Brillouin oscillators~\cite{Li2013a}, or electro-optical dividers~\cite{Li2014}, hold promise for their ability to synthesise low-noise or widely tunable microwave signals with compact form factor.
An additional approach is based on optical frequency division, which makes use of a self-referenced fs-laser comb optically-locked to an ultra-stable laser (USL) with a typical linewidth at the Hz-level~\cite{Millo2009a,Fortier2011,Portuondo-Campa2015,Xie2017}. If the comb line of index $N$ is tightly phase-locked to the USL (after subtraction of the carrier envelope offset (CEO) frequency $f_{\textsc{ceo}}$ or simultaneous stabilisation of $f_{\textsc{ceo}}$), the comb repetition rate $f_{\rm rep}$ is directly phase-stabilised to the ultra-stable frequency $\nu_{\textsc{usl}}$ by frequency division: $f_{\rm rep} = \nu_{\textsc{usl}}/N$. Importantly, owing to the carrier frequency division from optics to microwaves, the absolute phase noise power spectral density is reduced by a factor $N^2 \sim 10^8$.

This method has been mostly implemented using fibre-based fs-lasers with repetition rates of a few hundred megahertz. A fast actuator (e.g., an intra-cavity electro-optic modulator~\cite{Hudson2005}) is required to achieve a tight optical lock of the comb tooth to the optical reference and perform the frequency division over a wide bandwidth. Moreover, a high harmonic of the comb repetition rate must be used to synthesise a microwave signal beyond 10~GHz. Consequently, repetition rate multipliers are typically employed to reduce the impact of shot-noise in the photo-detection of the pulse train, such as optical filtering cavities~\cite{Diddams2009} or fibre interleavers~\cite{Haboucha2011}, which increases the system complexity. Therefore, the use of frequency combs directly operating at $\sim10$~GHz repetition rates would be highly beneficial, but their optical lock and self-referencing are challenging.

Microresonator-based Kerr frequency combs (i.e., `microcombs'), which naturally produce multi-GHz comb spectra generated via four-wave mixing in an optical microresonator~\cite{Gaeta2019,Kippenberg2018}, are natural candidate in this context. Pumping a cavity resonance with a continuous-wave laser can initiate and sustain a circulating dissipative Kerr soliton (DKS) pulse~\cite{Herr2013,Yi2015,Brasch2015,Joshi2016} that is intrinsically phase-coherent with the input pump laser. The resulting comb coupled out of the micro-cavity is inherently perfectly phase-locked to the pump laser, without any actuator locking bandwidth limitation. Direct soliton generation from an ultra-stable pump laser holds potential for compact and powerful optical-to-microwave dividers. Although self-referenced optical microcombs and clocks have been demonstrated~\cite{Jost2015,DelHaye2016,Newman2018}, optical frequency division for ultralow-noise microwave generation using such devices has not been demonstrated so far, mainly due to the complex crosstalk occurring between their two degrees of freedom~\cite{DelHaye2016, Stone2018} and the limited performance of the available actuators~\cite{Papp2013a, Joshi2016}.

Here, we demonstrate the generation of an ultralow-noise microwave signal using a microcomb-based transfer oscillator method to realise optical-to-microwave frequency division. The transfer oscillator method~\cite{Telle2002,Brochard2018} bypasses the need for tight optical phase-locking of the frequency comb to the optical reference. Instead, it relies on an adequate manipulation and combination of signals to cancel the comb phase noise and to provide a broadband electronic division of the USL frequency to the microwave domain. The frequency division by a large factor $N$ is performed electronically, thus removing the need for high locking bandwidth actuators. In this work, the USL is used to pump the microresonator and inherently constitutes a tooth of the resulting frequency comb. We show how to extend the transfer oscillator technique to exploit this salient feature of microcombs (or equivalently of electro-optic combs~\cite{Beha2017}).
In this proof-of-principle demonstration, we achieved a measured single-sideband phase noise of $-110$~dBc/Hz at 200~Hz offset from the 14.09~GHz carrier, which is 15~dB below the lowest phase noise microresonator-based photonic oscillator reported so far~\cite{Liang2015}, demonstrating the potential of this approach.

\section*{Results}
\subsection*{Transfer oscillator principle}
The high-level working principle of our method is illustrated in \fref{fig:concept}. A microresonator pumped by a sub-Hz-linewidth USL at frequency $\nu_{\textsc{usl}}$ generates a soliton-Kerr comb with a GHz-range repetition rate $f_{\rm rep}$ that is set by the resonator free spectral range (FSR). The reference laser is part of the frequency comb (line $N$) such that its frequency can be written as $\nu_{\textsc{usl}} = f_{\textsc{ceo}} + N \, f_{\rm rep}$. The detection of the CEO frequency (for example via $f-2f$ interferometry~\cite{Telle1999,Jones2000} or with an auxiliary self-referenced comb as in the present work) is followed by electronic division by means of a combination of frequency pre-scalers and direct digital synthesisers (DDS). The final step consists of mixing the divided CEO signal with the repetition rate, which yields 
\begin{equation}
f_{\rm signal} = \frac{f_{\textsc{ceo}}}{N} + f_{\rm rep} =  \frac{\nu_{\textsc{usl}}}{N}
\end{equation}
Importantly, this process can be carried out with a free-running Kerr comb and circumvents the need for a high-bandwidth repetition rate lock.

\subsection*{Microcomb generation with the USL}
Due to the limited tuning of the resonator and narrow bandwidth of the microcomb used for this proof-of-concept, additional hardware was needed to demonstrate the transfer oscillator principle, as shown in \fref{fig:setup:setup}.
The soliton-Kerr comb is generated by pumping a crystalline magnesium fluoride (\ce{MgF2}) microresonator with an FSR of 14.09~GHz using a 1553~nm diode laser, which is initially quickly scanned across a resonance for soliton generation~\cite{Herr2013}. Next, the pump laser is phase-locked to a sub-Hz-linewidth USL (Menlo Systems ORS1500) at a frequency detuning of $\sim 1.7$~GHz, thereby acquiring a comparable level of purity and stability (green box in \fref{fig:setup:setup}, a detailed description is provided in the \SIref{SI:PLL_soliton}).

To ensure the long-term stable operation of the Kerr comb and prevent the decay of the soliton, the microresonator resonance is slowly locked to the pump laser with an effective-detuning stabilisation, achieved via a sideband Pound-Drever-Hall (PDH) lock~\cite{Thorpe2008b}, which feedbacks on an acousto-optic modulator (AOM) that modulates the pump power and thus thermo-optically tunes the resonator, in addition to a slow thermal actuation of the microresonator (see details in the \SIref{SI:res_stab}). The detuning setpoint was carefully optimised in order to minimise the noise of the Kerr comb repetition rate $\fkrep$ at offset frequencies beyond $\sim 100$~Hz (see the Methods section and \fref{fig:KFC_RIN:PN}). However, the residual thermal drift of the resonator degrades the performance at lower offset frequencies.

\subsection*{Offset detection with an auxiliary comb}
The used crystalline \ce{MgF2} micro-comb features a relatively narrow spectrum that prevents a direct detection of its CEO frequency (\fref{fig:setup:Optspectrum}). The self-referencing of Kerr combs remains highly demanding due to the high repetition rate, low optical power, and fairly long pulse duration (225~fs here) resulting in a low peak intensity that makes the spectral broadening for $f-2f$ interferometry challenging~\cite{DelHaye2016,Lamb2018}.
Therefore, we implemented an indirect detection scheme using an auxiliary self-referenced fibre-laser frequency comb~\cite{Tian2018a} with a repetition rate $\fauxrep = 251.7$~MHz, as schematised in \fref{fig:setup:injlock}. Provided the repetition rates of both combs are harmonically phase-locked, i.e., $\fkrep = M \, \fauxrep$ (superscripts `K' and `aux'  refer to the Kerr and auxiliary combs, respectively), then the optical beatnote between the two combs corresponds to their relative CEO frequency $\Delta f_{\textsc{ceo}} = \fkceo - \fauxceo$, as the repetition rate noise contributions compensate each other in this beat signal. The Kerr comb CEO frequency is then obtained by mixing out the CEO frequency of the auxiliary comb $\fauxceo$ detected with an $f-2f$ interferometer (see \fref{fig:setup:setup}) and corresponds to $\fkceo = \Delta f_{\textsc{ceo}} + \fauxceo = \nu_{\text{pump}} - N \fkrep$ (grey box in \fref{fig:setup:setup}). Importantly, the auxiliary comb is not stabilised to the USL at any point and thus does not perform the division. Its role is limited to the offset detection in this demonstration.

The mutual harmonic phase-locking of the comb repetition rates is achieved via soliton injection-locking~\cite{Weng2019}. The harmonic $M=56$ of the repetition rate of the auxiliary comb (at 14.093~GHz) is detected, filtered and amplified to phase-modulate the pump light using an electro-optic modulator (EOM, blue box in \fref{fig:setup:setup}). This frequency is very close to the native microcomb line spacing, which gets injection-locked to this drive signal. Therefore, both repetition rates are strongly correlated over a bandwidth of $\sim 2$~kHz (see the \SIref{SI:InjLocking} and Supplementary Figure~\ref{fig:INJLOCK}).

\subsection*{Transfer oscillator chain}
The Kerr comb CEO signal, indirectly obtained as previously described, is detected at low frequency (MHz-range) and filtered to match the bandwidth of the injection locking of the repetition rate (not represented in \fref{fig:setup:setup}, see the \SIref{SI:TO_chain} and Supplementary Figure~\ref{fig:DivisionChain}). After up-mixing to 15~GHz, it is frequency-divided by a large pre-determined factor $N\approx 13,698$ and is subtracted to the separately-detected repetition rate $\fkrep$ to obtain the frequency-divided signal of the ultra-stable pump laser: $\nu_{\text{pump}}/N = \fkceo/N + \fkrep$ (orange box in \fref{fig:setup:setup}). The overall division of the Kerr comb CEO signal by the factor $N$ is realised with a frequency pre-scaler followed by two parallel DDS, which offers improved filtering capabilities in the electronic division~\cite{Brochard2018}. This second stage division with the DDS allows for a precise non-integer frequency division factor and leads to a clean single-tone output signal corresponding to the frequency-divided USL (see \fref{fig:result:Fine_RF_Spectr}).
The detailed description of the frequency division chain is provided in the \SIref{SI:TO_chain}.

The overall division factor $N$ was accurately determined experimentally, without prior knowledge of the optical frequency of the ultra-stable pump laser, by measuring the frequency change of the generated microwave signal corresponding to a small variation (140~Hz) of the Kerr comb repetition rate for different programmed division factors $N$ (see \fref{fig:result:ratio}). This simple measurement also provides an accurate determination of an optical comb line index $N$ that can be useful for absolute optical frequency measurements.

\subsection*{Microwave characterisation}
The phase noise of the generated ultralow-noise 14.09~GHz signal was measured with a cross-correlator phase noise analyser (\fref{fig:result:PN}). It reaches $-110$~dBc/Hz at 200~Hz Fourier frequency, 15~dB below the lowest phase noise microresonator-based photonics oscillator~\cite{Liang2015} at 10~GHz. The phase noise is below $-135$~dBc/Hz at 10~kHz and $-150$~dBc/Hz at around 1~MHz, showing that the intrinsic low-noise properties of the soliton Kerr comb at high Fourier frequencies are preserved. The calculated shot-noise predicts a noise floor at $-152$~dBc/Hz (thermal noise floor $\sim -170$~dBc/Hz). At 1 Hz offset, the measurement is limited by the instrumental noise floor below $ -60 $~dBc/Hz, even with 3000 cross correlations. Nevertheless, the transfer oscillator offers an improvement by at least 40~dB compared to the direct detection of the Kerr comb repetition rate (despite the resonator being stabilised to the USL), showing its ability to cancel the residual thermal drifts of the Kerr cavity.
The technical limitations of the measurement, at low offset frequency, make it difficult to directly compare the method with state-of-the-art optical frequency-division using mode-locked lasers. Nevertheless, at high Fourier frequencies, our results surpass some of the first demonstrations of optical frequency division~\cite{Millo2009a}, even if no optimisation has been performed on the photodetection side. Over the past 10 years, the development of mode-locked lasers, as well as the improvement of photodetection noise~\cite{Ivanov2005,Bouchand2017}, led to a reduction of the noise of the generated microwaves by 30 to 40~dB in some frequency bands~\cite{Xie2017}. We believe that the transfer oscillator method can follow a similar path, as in particular, the high repetition rates of the Kerr combs should make the photodetection optimisation less stringent.

\section*{Discussion}
In summary, we have reported optical-to-microwave frequency division using a Kerr comb as transfer oscillator. This demonstrates the potential of this method in microwave photonics and enlarges its previously reported implementation with low repetition rate mode-locked lasers. The approach presented here can be further implemented with electro-optic combs, where self-referencing and feedback control were recently achieved~\cite{Carlson2018,Beha2017}.
Although this proof-of-principle experiment required an auxiliary comb to obtain the CEO frequency of the Kerr comb, directly self-referenced microcombs are technologically feasible in silicon nitride (\ce{Si3N4}) photonic-chips~\cite{Brasch2017}.
While octave-spanning comb spectra have been achieved using dispersion control~\cite{Pfeiffer2017,Newman2018}, these implementations used THz repetition rates to cover such a large spectral range, which made photodetection of the repetition rate practically impossible. Nonetheless, the residual phase noise of these combs has been shown to be suitable for frequency division~\cite{Drake2019a}.
Recent improvements of integrated resonators have enabled soliton microcombs with K- and X-band (20 and 10~GHz) repetition rates in integrated resonators~\cite{Liu2019}. However, the achieved spectral spans, although wider than in the crystalline case, are far from covering one octave. Pulsed  pumping~\cite{Obrzud2017} appears as a promising approach to enable octave spanning microcombs with detectable microwave repetition rates.
This approach uses synchronous pumping of the microresonator with picosecond pulses to generate a soliton with a much shorter duration and a spectrum that can cover an octave, similar to enhancement cavities~\cite{Lilienfein2019}. It can be seen as a hybrid between an electro-optic (EO) comb and a microcomb, with the advantage that the spectral enlargement of the EO comb is performed in cavity and is therefore directly filtered~\cite{Anderson2019,Brasch2019}.
Crucially, even if the free-running phase noise of these integrated microcombs is typically higher than in the crystalline platform used in this work~\cite{Liu2019,Huang2019,Drake2019}, the additional noise is cancelled over a broad frequency range via the transfer oscillator method that constitutes a powerful tool for low-noise frequency division without the need for a very low-noise comb.
The free-running comb operation and the maturity of RF components, which can be suitably integrated, promise robust device operation.
Furthermore, improvements in resonator actuation, using micro-heaters~\cite{Joshi2016}, piezoelectric transducers~\cite{Liang2017a,Tian2018,Dong2018}, or the electro-optic effect~\cite{Alexander2018,He2019}, will allow the resonator to be tuned to the USL for direct soliton generation (as in \fref{fig:concept:mockup}), alleviating the need for an optical phase-lock loop and greatly simplifying the detuning stabilisation mechanism.
If a lower stability level is acceptable, simpler and more compact low-noise lasers can be employed~\cite{Kefelian2009,Liang2015a,Gundavarapu2019} instead of the USL.
We believe that the presented transfer oscillator method holds promising potential for ultralow-noise high-frequency generators with a new generation of compact photonic-based systems~\cite{Marpaung2019} for radar applications~\cite{Ghelfi2014}, high frequency telecommunications~\cite{Koenig2013} and time–frequency metrology~\cite{Millo2009a}.


\section*{Methods}
\begin{figure*}[!t]
	\centering
	\includegraphics[width=\textwidth]{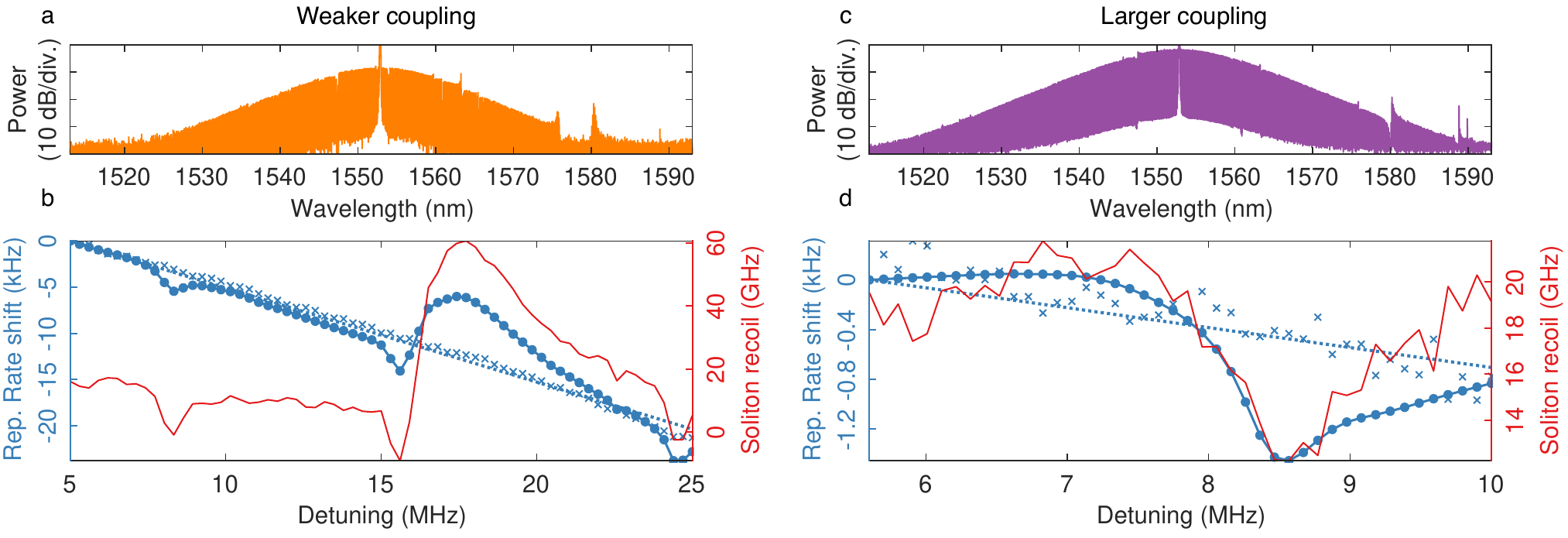}
	{\phantomsubcaption\label{fig:KFC_optim:spectrUndc}} 
	{\phantomsubcaption\label{fig:KFC_optim:rep_redunUndc}} 
	{\phantomsubcaption\label{fig:KFC_optim:spectrOvc}} 
	{\phantomsubcaption\label{fig:KFC_optim:rep_redunOvc}} 
	\caption{\textbf{\boldmath Optimisation of $\fkrep$ phase noise}
		\subref{fig:KFC_optim:spectrUndc} 
		Soliton spectrum for lower coupling case (Detuning 10~MHz).
		\subref{fig:KFC_optim:rep_redunUndc} 
		Evolution of the repetition rate (blue, solid) and of the soliton recoil ($\Omega/2\pi$) retrieved by fitting the optical spectrum (red), in the lower coupling case. The blue crosses and dashed line show the residual repetition rate change after subtraction of the recoil induced shift (using eq.~\eqref{eq:shiftRepRate}).
		\subref{fig:KFC_optim:spectrOvc} 
		Soliton spectrum for larger coupling case (Detuning 10~MHz).
		\subref{fig:KFC_optim:rep_redunOvc} 
		Evolution of the repetition rate (blue, solid) and of the soliton recoil ($\Omega/2\pi$) retrieved by fitting the optical spectrum (red), in the larger coupling case. The blue dashed line shows the residual repetition rate change after subtraction of the recoil induced shift (using eq.~\eqref{eq:shiftRepRate}).
	}
	\label{fig:KFC_optim}
\end{figure*}

\subsection*{Operating conditions}
The pump power after the EOM used for the PDH lock of the microresonator and the injection locking of $f_{\rm rep}$ is $\sim 10$~mW and is amplified to $\sim 250$~mW in an EDFA. The power level after the AOM that controls the pump power coupled to the resonator (see \fref{fig:setup:setup}) is set to $\sim 210$~mW. After comb generation and residual pump rejection with a fibre Bragg grating, the comb power of $\sim 1$~mW is amplified to $\gtrsim 5$~mW. The largest part of this power (90\%) is sent onto a high power handling photodiode (Discovery Semiconductors DSC40S, generating a photocurrent of 5.12~mA and a microwave power of $\sim -7.4$~dBm), while the remaining fraction is used for the intercomb beatnote detection. The shot noise level is estimated for a CW laser detection, based on the generated photocurrent and microwave power.
The 56\textsuperscript{th} harmonic of the auxiliary comb repetition rate $\fauxrep$ at 14.09~GHz is detected, selected using a narrow band-pass filter and amplified to $\sim 19$~dBm. This signal drives the phase modulator and creates an estimated phase deviation of $\sim 1.4$~rad. The injection-locking range of the Kerr comb repetition rate~\cite{Weng2019} spans $\gtrsim 2$~kHz and the locking bandwidth is $\sim 2$~kHz.

\subsection*{Resonator characteristics}
The \ce{MgF2} whispering gallery mode resonator was fabricated via precision diamond turning and hand polishing on a lathe. The intrinsic linewidth of the pumped mode is $\sim 80$~kHz (intrinsic quality factor of $2.4 \times 10^{9}$). The evanescent coupling to the resonance is achieved via a tapered optical fibre. The fibre is operated in contact with the resonator to damp its vibrations. Careful adjustment of the fibre position is required to maximise the coupling rate and increase the out-coupled comb power. The loaded resonance linewidth is estimated at $\sim 2.4$~MHz. The threshold power for comb formation is estimated at $ \sim 40 $~mW. The detuning setpoint was chosen to minimise the noise of the Kerr comb repetition rate, as described in the next section.

\subsection*{Soliton noise minimisation}
\label{sec:noiseoptim}
The laser-resonator detuning $\delta = \nu_{\rm cav} - \nu_{\rm laser}$ is known to have a major impact on the noise and stability of Kerr frequency combs. This parameter not only sets the soliton pulse duration~\cite{Lucas2017}, but was also shown to modify the repetition rate frequency through the Raman self-frequency shift~\cite{Yi2016a} $\Omega_{\rm Raman}(\delta)$ and the soliton recoil $\Omega_{\rm recoil}$ corresponding to dispersive wave emission~\cite{Yi2016b,Lucas2017}.
Indeed, these two effects lead to an overall shift of the spectral centre of the soliton (i.e., the soliton spectral maximum relative to the pump frequency) $\Omega = \Omega_{\rm Raman} + \Omega_{\rm recoil}$, which induces in turn a change in the group velocity of the pulse and therefore of the repetition rate according to~\cite{Matsko2013}
\begin{equation}
\fkrep = \dfrac{1}{2\pi} \left( D_1 + \dfrac{D_2}{D_1} \, \Omega(\delta) \right) \label{eq:shiftRepRate}
\end{equation}
where $D_1/2\pi = 14.09$~GHz is the resonator FSR and $D_2/2\pi = 1.96$~kHz is the group velocity dispersion (GVD) parameter at the pump frequency~\cite{Herr2012}.
Thus, residual laser-resonator detuning noise can degrade the spectral purity of the repetition rate~\cite{Stone2018}. A solution to this problem was already identified by Yi et al.~\cite{Yi2016b}, who proposed to use the balance of dispersive-wave recoil and Raman-induced soliton-self-frequency shift to enhance the repetition-rate stability of a silica wedge-based Kerr comb. A similar concept is applied here to minimise the repetition rate noise of the crystalline \ce{MgF2} microresonator-based comb. Importantly, in \ce{MgF2}, the Raman self frequency shift can be neglected, due to the very narrow gain bandwidth, and the soliton shift is dominated by the soliton recoil $\Omega \approx \Omega_{\rm recoil}$.

We measured the variation in repetition rate of the soliton comb as a function of detuning in two coupling conditions (weak and large coupling). The coupling rate was modified by changing the position of the tapered fibre along the resonator. The detuning was scanned (forward and backward) by changing the PDH modulation frequency, while the phase lock loop offset frequency was adapted accordingly to keep the total frequency offset between the USL and the microresonator resonance constant. At each detuning point, the optical spectrum was acquired and the repetition rate frequency $\fkrep$ was counted. The results are displayed in \fref{fig:KFC_optim}. The phase modulation at the cavity FSR used for injection-locking was disabled in this measurement.

The weak coupling of the resonator allows for a relatively wide detuning range to be accessed (5 to 25~MHz, see \fref{fig:KFC_optim:rep_redunUndc}). Over this span, the repetition rate changes in total by 22~kHz, but not linearly. The non-monotonic evolution of $\fkrep (\delta)$ is caused by the soliton recoil induced by dispersive waves through avoided mode crossings~\cite{Yang2016c,Yi2016b,Lucas2017}. The soliton shift $\Omega/2\pi$ is extracted by fitting the optical spectrum with a $\operatorname{sech}^2$ function and the associated repetition rate variation can be estimated using eq.~\eqref{eq:shiftRepRate}. Interestingly, after subtracting this contribution, the residual shift of the repetition rate follows a linear trend with a slope of $\sim -1$~kHz/MHz. This significant variation is independent from any recoil-associated effect and could originate from more complex forms of avoided modal crossings, or third order dispersion, although we observed that the value of this slope changes with the coupling as detailed below.

Increasing the coupling rate of the resonator (see \fref{fig:KFC_optim:rep_redunOvc}) shrinks the accessible detuning range (5.5 to 10~MHz) and radically changes the dependence of $\fkrep$ with $\delta$. The overall variation is reduced to $\sim 1.4$~kHz and is dominated by solitonic recoil. Once this contribution is subtracted, the residual slope is on the order of $\sim -160$~Hz/MHz, which is very close to the value expected from the nonlinear self-steepening effect~\cite{Bao2017c}.

\begin{figure}[t]
\centering
\includegraphics[width=\columnwidth]{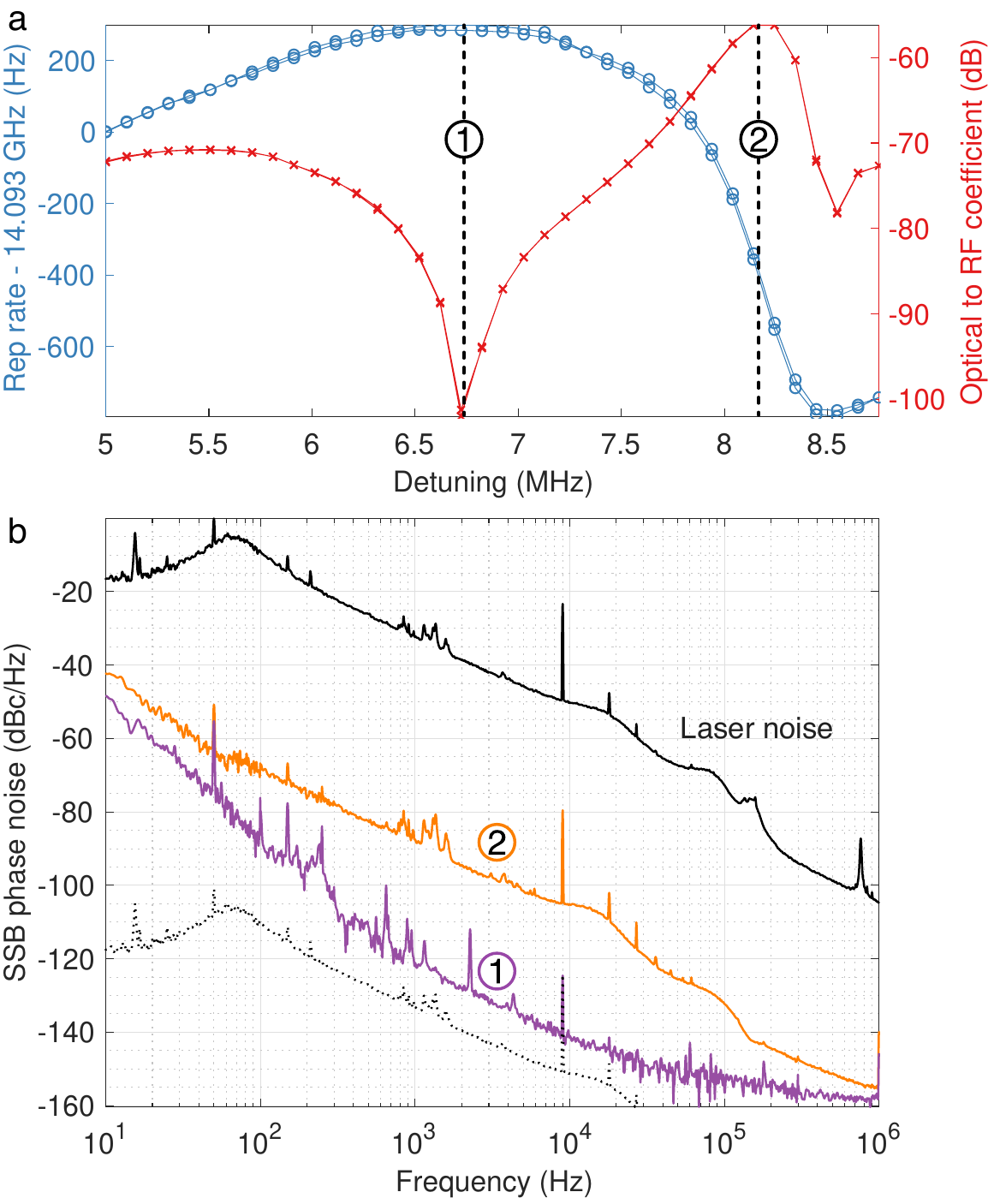}
{\phantomsubcaption\label{fig:KFC_optim_PN:Scan}}
{\phantomsubcaption\label{fig:KFC_optim_PN:PN}}
\caption{\textbf{`Quiet' operating point}
\subref{fig:KFC_optim_PN:Scan} Evolution of the repetition rate with the detuning (blue) and associated optical phase modulation to RF phase modulation conversion coefficient calibrated with the 9~kHz phase modulation tone on the laser (red).
\subref{fig:KFC_optim_PN:PN} Phase noise spectra of the soliton repetition rate at the two operating points highlighted in (a). The solid black line shows the laser noise (PDH-stabilised to the microcavity). The dashed black line shows the noise of the laser scaled by $-100$~dB to match the 9~kHz phase calibration tone.
}
\label{fig:KFC_optim_PN}
\end{figure}

\subsection*{Quiet operation point}
More notably, under this larger coupling condition, the relation $\fkrep (\delta)$ exhibits a stationary point around $\delta = 7$~MHz, where the coupling of pump-laser frequency noise into the soliton repetition rate is expected to be minimal since  $\partial \fkrep / \partial \delta \approx 0$. To verify this prediction, the phase noise of the detected soliton pulse train was measured at different detuning points. The pump laser was phase-modulated by a low frequency tone at 9~kHz to provide a reference point. Furthermore, instead of phase-locking the pump laser to the USL, the PDH feedback was applied to the pump laser current in these measurements, and the resonator was slowly stabilised to the USL via power and thermal feedback. The larger laser noise obtained in this case helps visualising its impact on the repetition rate frequency and could be calibrated via a heterodyne measurement with the USL. The results are displayed in \fref{fig:KFC_optim_PN}. At the operating point~2, where the slope of $\fkrep (\delta)$ is maximum, the noise of $\fkrep$ follows the same features as the laser noise. Rescaling the laser noise to match the 9 kHz modulation peaks indicates that the optical noise is reduced by $56$~dB. Conversely the point~1, where the slope of $\fkrep (\delta)$ is minimum, corresponds to the lowest optical-to-RF noise transduction (dip in \fref{fig:KFC_optim_PN:Scan}), with a conversion coefficient below $-100$~dB.  As expected, this point yields the lowest achieved phase noise, and it appears that the laser phase noise is no longer the overall limiting factor of the Kerr comb repetition rate noise.

\begin{figure}[t]
\centering
\includegraphics[width=\columnwidth]{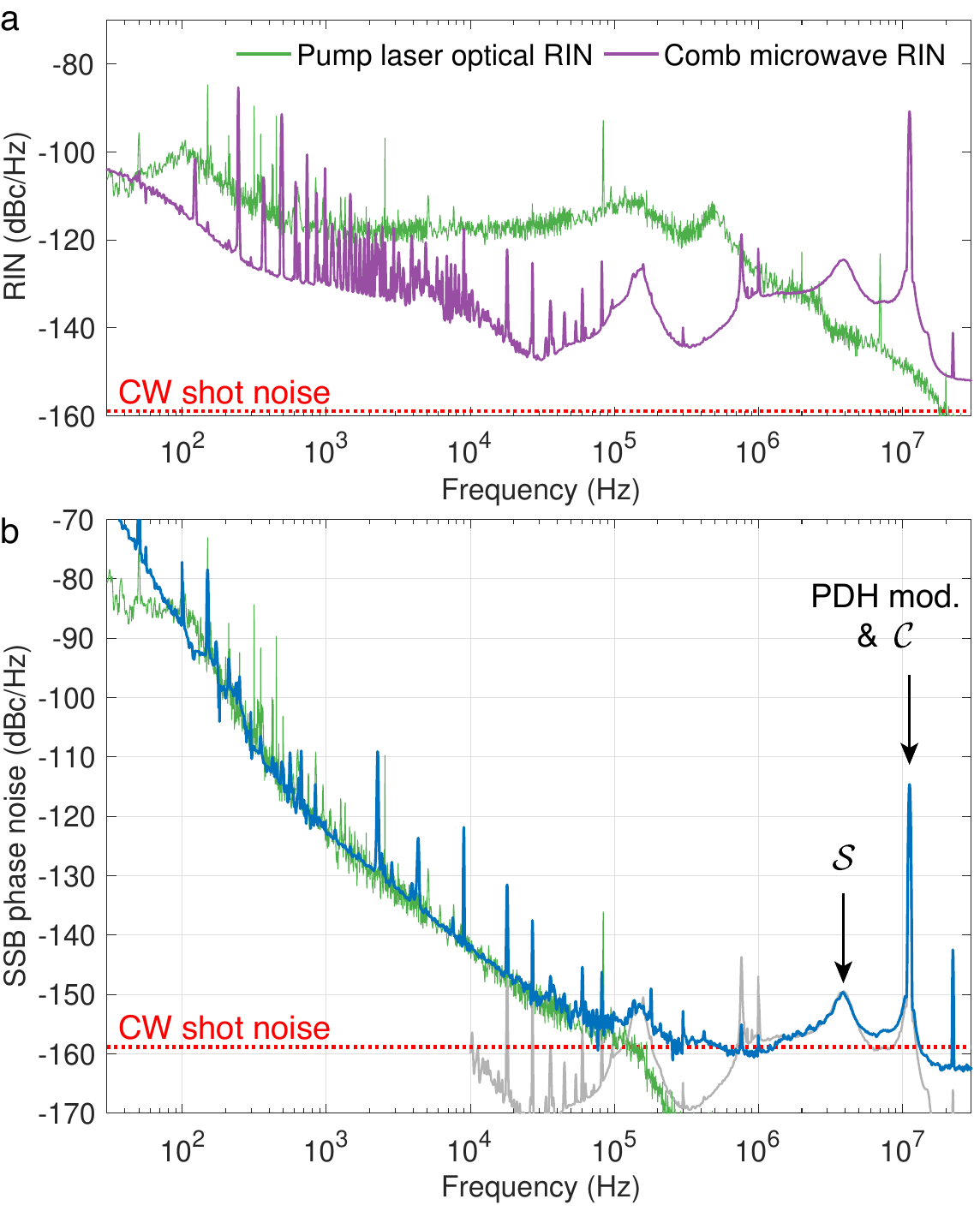}
{\phantomsubcaption\label{fig:KFC_RIN:RIN}}
{\phantomsubcaption\label{fig:KFC_RIN:PN}}
\caption{\textbf{Pump laser RIN and estimated limitation on the phase noise}
\subref{fig:KFC_RIN:RIN} Optical RIN of the pump laser (green) and microwave amplitude noise of the soliton repetition rate (purple).
\subref{fig:KFC_RIN:PN} Phase noise spectrum of the repetition rate in the `quiet' point (blue) and estimated limitation from the pump laser RIN (green). The grey curve corresponds to the estimated AM-to-PM conversion in the photodiode (microwave amplitude noise scaled by $-25$~dB).
}
\label{fig:KFC_RIN}
\end{figure}

\subsection*{Noise limitations in microcombs}
In a nonlinear resonator, the free spectral range $D_1/2\pi$ depends on the circulating optical power. Therefore, the relative intensity noise (RIN) of the pump laser (power $P_{\rm in}$) eventually induces timing jitter of the repetition rate, according to eq.\eqref{eq:shiftRepRate}. Assuming a laser on resonance, the self phase modulation induced shift follows~\cite{Wilson2018}:
\begin{equation}
\dfrac{\delta D_1(\omega)}{2\pi} = \underbrace{ \left( \dfrac{D_1}{2\pi} \dfrac{4 \eta \, c \, n_2}{\kappa V_{\rm eff} n_0^2} \right) }_{\alpha} \, \delta P_{\rm in}(\omega)
\end{equation}
where $\kappa/2\pi \approx 1.35$~MHz is the cavity energy decay rate, $\eta = \kappa_{\rm ex} / \kappa \approx 0.94$ is the coupling impedance of the resonator ($\kappa_{\rm ex}$ is the coupling rate), $V_{\rm eff} \approx 2.32\times 10^{-12}~\meter^3$ is the mode volume, $n_2 = 9 \times 10^{-21}~\meter^2\per\watt$ is the (Kerr) nonlinear index and $n_0 = 1.37$ is the refractive index. These values yield a conversion coefficient $\alpha \approx 3.8~\kilo\hertz\per\watt$. We measured the relative intensity noise $S_{\rm RIN}(f)$ of the pump laser (\fref{fig:KFC_RIN:RIN}) and the associated induced phase noise was estimated using:
\begin{equation}
S_{D_1/2\pi}^{\phi} (f) = \left( \dfrac{\alpha}{f} P_{\rm in} \right)^2 S_{\rm RIN}(f)
\end{equation}
for the measured input pump power of $P_{\rm in} \approx 212$~mW. The results are displayed in \fref{fig:KFC_RIN:PN}. The estimated level matches remarkably the repetition rate phase noise at offsets between 500~Hz and 100~kHz (blue and green curves in \fref{fig:KFC_RIN:PN}), suggesting that the pump laser RIN is limiting the performances in this range. The phase noise reaches $\sim -143$~dBc/Hz at 10~kHz, which outperforms any other microresonator-based approach~\cite{Liang2015,Li2014,Li2013a,Yi2016b,Liang2017a}.

At lower offset frequencies (50 -- 500 Hz), the thermal fluctuations and drift of the resonator, which are beyond the power stabilisation bandwidth, are the limiting factor~\cite{Gorodetsky2004,Liang2015}.

At higher offset frequencies, two noise bumps appear related to the characteristic double resonant response ($\mathcal{S}$ and $\mathcal{C}$) of the resonator in the soliton regime~\cite{Guo2016}. In these resonant features, the transduction of the pump laser noise is enhanced~\cite{Stone2018}. Beyond 100~kHz offset, the contributions of various factors are more difficult to identify. We observed nonetheless a correlation between the microwave RIN (\fref{fig:KFC_RIN:RIN}) and the phase noise, which suggests that amplitude-to-phase noise conversion is occurring in the photodiode~\cite{Zhang2011a}, with a conversion of $\sim -25$~dB (grey curve in \fref{fig:KFC_RIN:PN}), which is in agreement with reported values for similar photodiodes~\cite{Bouchand2017}. We report here the microwave amplitude noise, as our measurement device offered a better sensitivity in this configuration, but our observations showed that this amplitude noise matches well the optical RIN (measured at DC with a diplexer).

Finally, the continuous-wave shot-noise floor is expected to be at $-159$~dBc/Hz (photocurrent of 6.85~mA, microwave power of $-3.8$~dBm). However, we noticed that the phase noise floor of our measurement stands 4.1~dB below this value (at frequency offsets above 20~MHz), while the amplitude noise floor is 6.3~dB above. This imbalance between amplitude and phase needs further investigation and could be related to shot noise correlations in the detection of optical pulses~\cite{Niebauer1991,Quinlan2013a,Quinlan2013}.

\medskip
\section*{Data availability statement}
The data and code used to produce the results of this manuscript are available on Zenodo:\url{http://doi.org/10.5281/zenodo.3515211}.

\section*{Authors contributions}
E.L. and P.B. designed the experimental setup and performed the experiments with assistance of R.B. and S.S. E.L. analysed the data and wrote the manuscript, with input from other authors.  T.S. and T.J.K. supervised the project.

\section*{Competing interests}
The authors declare no competing interests.

\begin{acknowledgments}
The authors acknowledge Menlo Systems and the Observatoire de Paris (SYRTE) for providing access to the raw data of the measurement of the USL phase noise (the methods can be found in ref \cite{Giunta2018}). The authors thank M. H. Anderson for the proofreading the manuscript.
This publication was supported by the Swiss National Science Foundation (SNF) under grant agreement 176563, as well as Contract W31P4Q-14-C-0050 (PULSE) from the Defense Advanced Research Projects Agency (DARPA), Defense Sciences Office (DSO).
This material is based upon work supported by the Air Force Office of Scientific Research, Air Force Material Command, USAF under Award No. FA9550-15-1-0099. 
E.L. acknowledges support from the European Space Technology Centre, with ESA Contract No.~4000118777/16/NL/GM.
\end{acknowledgments}

\bigbreak
\def\bibsection{}  
\section*{References}
\medbreak
\bibliography{library}

\end{document}


\title{Supplementary Material for: Ultralow-Noise Photonic Microwave Synthesis using a Soliton Microcomb-based Transfer Oscillator}

\author{Erwan~Lucas}
\thanks{These authors contributed equally to this work.}
\affiliation{Institute of Physics, École Polytechnique Fédérale de Lausanne (EPFL), CH-1015 Lausanne, Switzerland}

\author{Pierre~Brochard}
\thanks{These authors contributed equally to this work.}
\affiliation{Laboratoire Temps-Fréquence, Université de Neuchâtel, CH-2000 Neuchâtel, Switzerland}

\author{Romain~Bouchand}
\affiliation{Institute of Physics, École Polytechnique Fédérale de Lausanne (EPFL), CH-1015 Lausanne, Switzerland}

\author{St\'ephane~Schilt}
\affiliation{Laboratoire Temps-Fréquence, Université de Neuchâtel, CH-2000 Neuchâtel, Switzerland}

\author{Thomas~Südmeyer}
\affiliation{Laboratoire Temps-Fréquence, Université de Neuchâtel, CH-2000 Neuchâtel, Switzerland}

\author{Tobias~J.~Kippenberg}
\affiliation{Institute of Physics, École Polytechnique Fédérale de Lausanne (EPFL), CH-1015 Lausanne, Switzerland}
\email{tobias.kippenberg@epfl.ch}

\maketitle

\def \fkrep {f_{\rm rep}^{\rm K}}
\def \fauxrep {f_{\rm rep}^{\rm aux}}
\def \fkceo {f_{\textsc{ceo}}^{\rm K}}
\def \fauxceo {f_{\textsc{ceo}}^{\rm aux}}
\def \fceo {f_{\textsc{ceo}}}
\def \frep {f_{\rm rep}}


\section{Soliton generation procedure and optical phase lock loop}
\label{SI:PLL_soliton}
In a microresonator with appropriate dispersion, solitons emerge spontaneously (soft excitation) when the continuous wave pump laser is scanned from blue (higher laser frequency) to red detuning (lower laser frequency) across a high-Q resonance~\cite{Herr2013}. This requires a rapid tunability of either the pump laser or the resonator. An ultra-stable laser (USL) locked to a high-finesse reference cavity is typically not tunable and the crystalline resonator used here cannot be tuned fast enough to overcome thermal effects~\cite{Carmon2004}. We circumvented this problem by implementing an optical phase lock loop (OPLL) to stabilise the frequency offset between the pump external-cavity diode laser (ECDL, Toptica CTL1550) and the reference USL (Menlo Systems ORS1500). The soliton state is generated by rapid tuning of the ECDL via current tuning. After soliton generation, the OPLL is activated with a frequency offset set so as to preserve the soliton. Resonators with improved tuning capabilities could remove the need for an OPLL altogether.

Solitons are sustained in the cavity within a narrow range of `red' laser--cavity detuning~\cite{Herr2013,Leo2010} (i.e., the laser frequency is lower than the resonance frequency), which depends on the pump power level and the coupling rate of the resonator. The detuning control is crucial in soliton-based Kerr comb as it determines many of the comb properties. First, in order to preserve the soliton in the cavity, the detuning must remain within the soliton existence range. Secondly, as exposed in the Methods section, the soliton properties, such as the pulse duration and especially the repetition rate, are highly sensitive to the detuning operating point~\cite{Yi2016b, Stone2018}. However, even if an USL is used to pump the resonator, the thermal drift of the resonator induces detuning drifts that can lead to degraded noise performance and even loss of the soliton. Therefore, following soliton generation and OPLL activation, an offset Pound-Drever-Hall (PDH) lock actively stabilises the detuning to a precise, defined radio-frequency (RF).

\begin{figure}[h]
\centering
\includegraphics[width=.8\columnwidth]{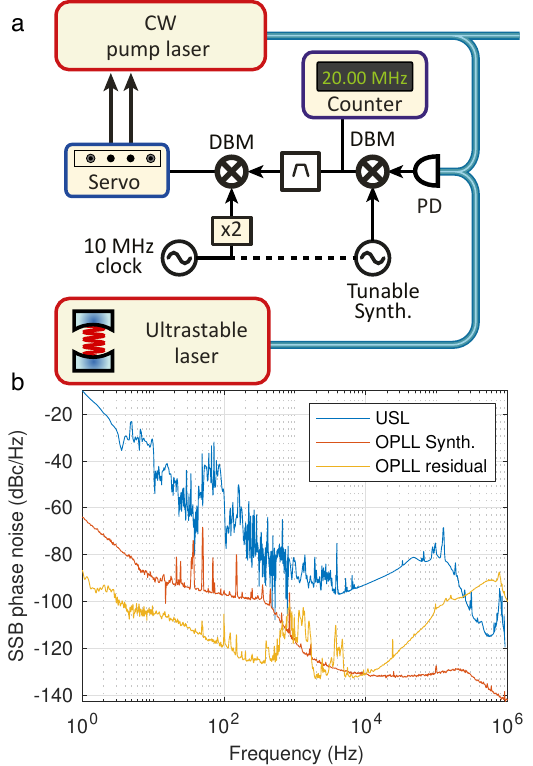}
{\phantomsubcaption\label{fig:OPLL:Setup}}
{\phantomsubcaption\label{fig:OPLL:BW}}
\caption{\textbf{Optical phase lock loop}
\subref{fig:OPLL:Setup} Detailed experimental setup. PD, photodiode; DBM double balanced mixer; $\times 2$, frequency doubler.
\subref{fig:OPLL:BW} Comparison between the USL phase noise and the residual phase noise contribution of the OPLL components. The red line shows the phase noise of the 1.7~GHz synthesiser and the yellow line the phase noise of the 20~MHz IF signal at the output of the DBM, indicating a feedback bandwidth in the range of $\sim 500$~kHz. The overall added noise is negligible compared to the USL.
}
\label{fig:OPLL}
\end{figure}

\begin{figure*}[t]
\centering
\includegraphics[width=.9\textwidth]{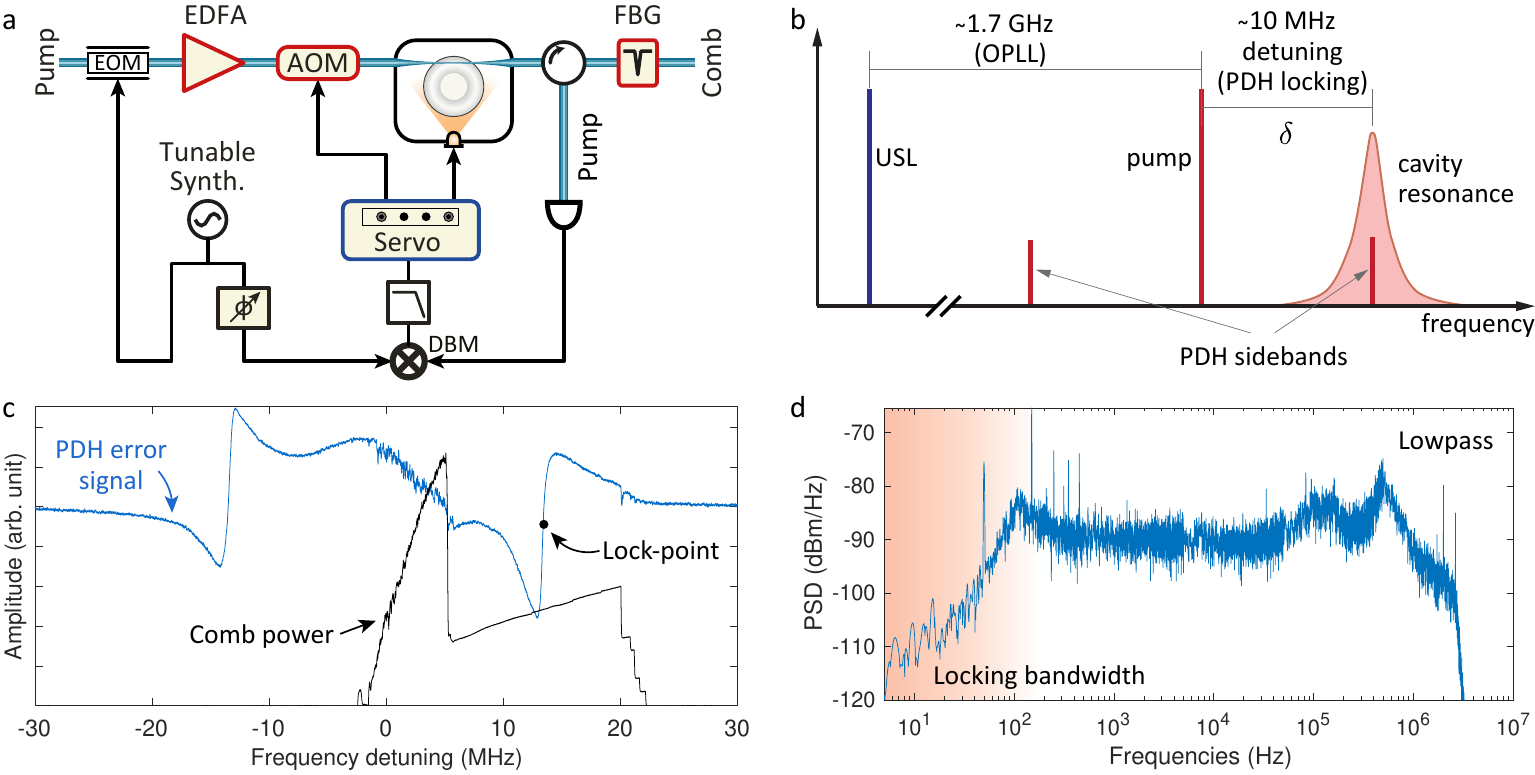}
{\phantomsubcaption\label{fig:PDH:Setup}}
{\phantomsubcaption\label{fig:PDH:Concept}}
{\phantomsubcaption\label{fig:PDH:Error}}
{\phantomsubcaption\label{fig:PDH:BW}}
\caption{\textbf{Pound-Drever-Hall detuning stabilisation}
\subref{fig:PDH:Setup} Detailed experimental setup. EOM, electro-optic modulator; EDFA, erbium-doped fibre amplifier; AOM, acousto-optic modulator; FBG, fibre Bragg grating.
\subref{fig:PDH:Concept} Scheme of principle of the stabilisation. The pump laser is phase-locked to the USL. The cavity resonance detuning is then locked to the pump laser via PDH stabilisation using a feedback to the pump laser power (and radiative heating of the resonator).
\subref{fig:PDH:Error} PDH error signal (blue) and generated comb power (black), as a function of detuning (the PDH frequency is 13.4~MHz). The detuning lock-point can be set arbitrary in the soliton step by changing the synthesiser frequency.
\subref{fig:PDH:BW} Power spectral density (PSD) of the residual PDH error signal, when the detuning lock is active.
}
\label{fig:PDH}
\end{figure*}

The schematic of the OPLL is presented in \fref{fig:OPLL:Setup}. The beatnote between the two lasers is photo-detected at a frequency of $1.7$~GHz, and down-mixed to an intermediate frequency (IF) of 20~MHz using a frequency synthesiser. After soliton generation, the synthesiser frequency is set precisely using a frequency counter, in order to preserve the pump laser within the soliton existence range upon lock activation. The IF signal is band-pass filtered and compared to a 20~MHz RF signal derived from a 10~MHz common clock, using a double-balanced mixer (DBM). The resulting error signal passes through a proportional–integral–derivative controller (PID, Toptica MFALC) that implements a slow feedback to the laser piezoelectric transducer and a fast feedback to the diode laser current allowing a $\sim 500$~kHz actuation bandwidth (see \fref{fig:OPLL:BW}). The residual noise of the OPLL and the phase noise of the synthesiser are negligible compared to the USL noise, indicating a good transfer of the USL purity to the pump laser (see \fref{fig:OPLL:BW}).

\section{Resonator stabilisation}
\label{SI:res_stab}
The detuning stabilisation is implemented via a PDH stabilisation (see \fref{fig:PDH}). The PDH error signal is obtained by phase-modulating the pump laser (at a frequency in the range of 5 -- 25~MHz) before coupling to the cavity, using an electro-optic modulator (EOM, iXblue MPX-LN-0.1). The modulated signal is detected after the resonator (on the filtered residual pump) and demodulated to DC using the same phase-shifted RF signal (to account for the unbalanced delay between the modulation and demodulation paths). In practice, a dual-channel arbitrary waveform generator is used and the relative phase between the channels is adapted as a function of the modulation frequency. After demodulation, the baseband signal is low-pass-filtered and sent to a PID servo-controller (Toptica FALC).
When the comb operates in the soliton regime, the pump laser is red-detuned. Thus, the PDH feedback setpoint corresponds to the higher frequency phase modulation sideband being in resonance (\fref{fig:PDH:Concept}).
The PDH servo acts thermally on the resonator, to maintain a fixed pump-resonator detuning. The feedback is implemented to the pump laser power (using a 0\textsuperscript{th} order acousto-optic modulator (AOM)) and a slower actuation on a LED (Thorlabs MCWHL5 with a typical power of 800 mW) shining on the resonator through a microscope also used for imaging. This allows keeping the pump power to a determined setpoint. The residual noise of the PDH error signal indicates an actuation bandwidth of $\sim 100$~Hz, limited by the thermal response of the resonator (\fref{fig:PDH:BW}).

Owing to this overall scheme, the resonator is stabilised to the USL, which improves the stability of the system and helps preserving a given operation point.


\section{Auxiliary comb and offset detection}

An auxiliary optical frequency comb (Menlo Systems FC1500) is used here to detect the CEO frequency of the Kerr-comb $\fkceo$, as the used crystalline micro-comb features a relatively narrow spectrum that prevents a direct detection of its CEO frequency. An optical beat-note between the two combs is first detected with a photodiode (NewFocus model 1811) at a frequency of a few tens of MHz. This low-frequency beat signal corresponds to the frequency difference between one mode of each comb, i.e.,
\begin{equation}
    f_{\rm beat} = N (\fkrep - 56 \, \fauxrep) + (\fkceo - \fauxceo) \label{eq:generalCombBeat}
\end{equation}
where the superscripts `K' and `aux' refer to the Kerr and auxiliary comb, respectively, and the 56\textsuperscript{th} harmonic of the 251.6~MHz repetition rate of the auxiliary comb is in close vicinity to the fundamental repetition rate of the Kerr comb.

\section{Injection locking}
\label{SI:InjLocking}
\begin{figure}[h]
\centering
\includegraphics[width=\columnwidth]{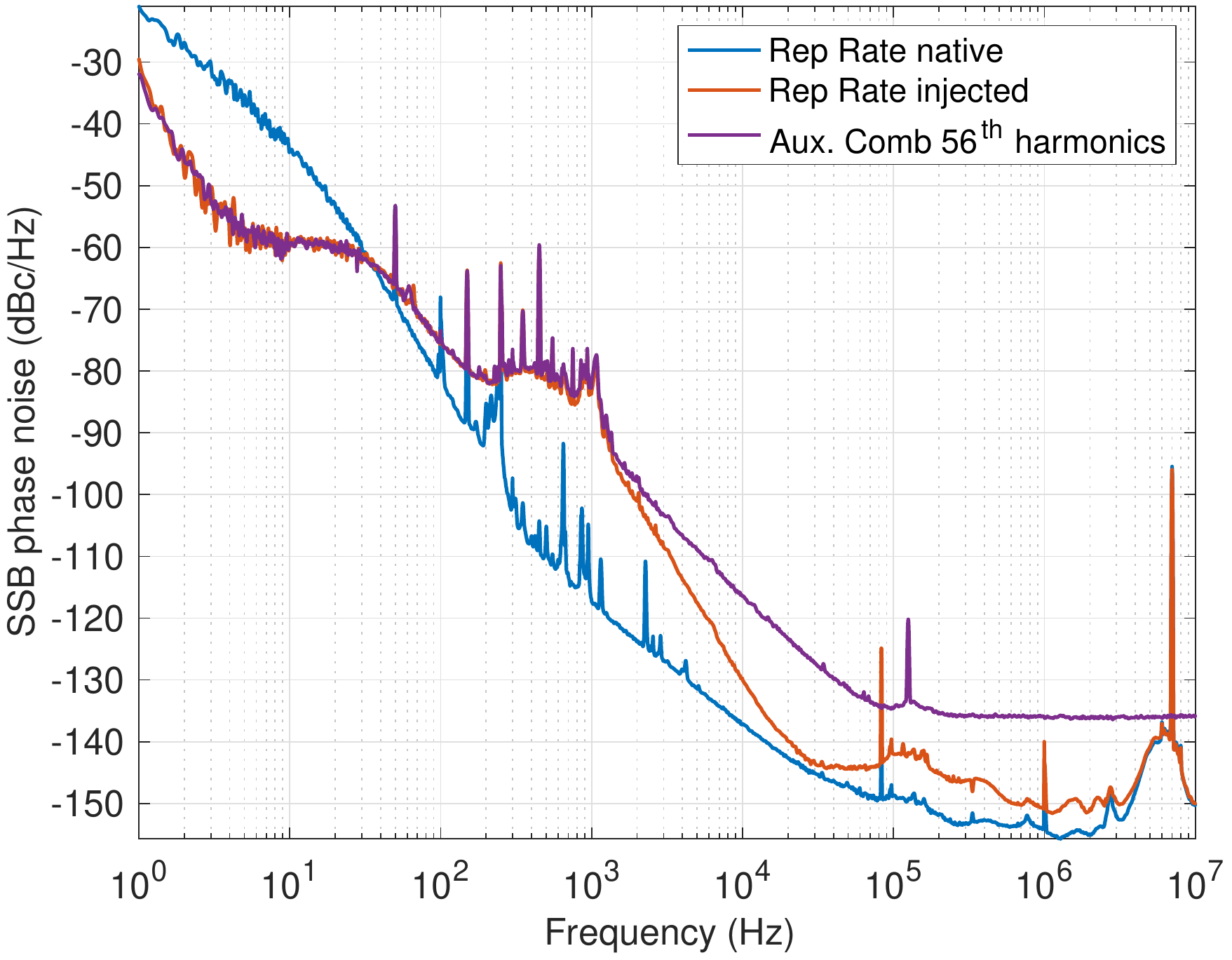}
\caption{\textbf{Injection locking of the repetition rate} Comparison between the phase noise of the Kerr comb repetition rate when it is native (blue) and injection locked (red) by the 56\textsuperscript{th} harmonic of the auxiliary comb repetition rate (purple).
}
\label{fig:INJLOCK}
\end{figure}

To suppress the relative phase noise between the repetition rate of the two combs in their beat signal, we imprint the $\frep$ noise of the auxiliary comb to the Kerr comb by injection locking. This is realised by detecting and band-pass filtering the 56\textsuperscript{th} harmonic of $\fauxrep$ (auxiliary comb) at 14.09 GHz and using this signal, after amplification to $\sim19$~dBm, to drive an EOM (iXblue MPZ-LN-10) to create a set of sidebands around the ultra-stable pump laser of the micro-resonator, which injection-lock the adjacent optical modes of the resonator. This strongly correlates the noise of the repetition rate of the two combs, but only within the bandwidth of the injection locking that is in the kHz range, as shown in \fref{fig:INJLOCK}. In that case, the beat signal frequency in eq. \eqref{eq:generalCombBeat} can be re-expressed as
\begin{equation}\label{key}
f_{\rm beat} = \fkceo - \fauxceo = \Delta \fceo
\end{equation}

\begin{figure*}
	\centering
	\includegraphics[width=.7\textwidth]{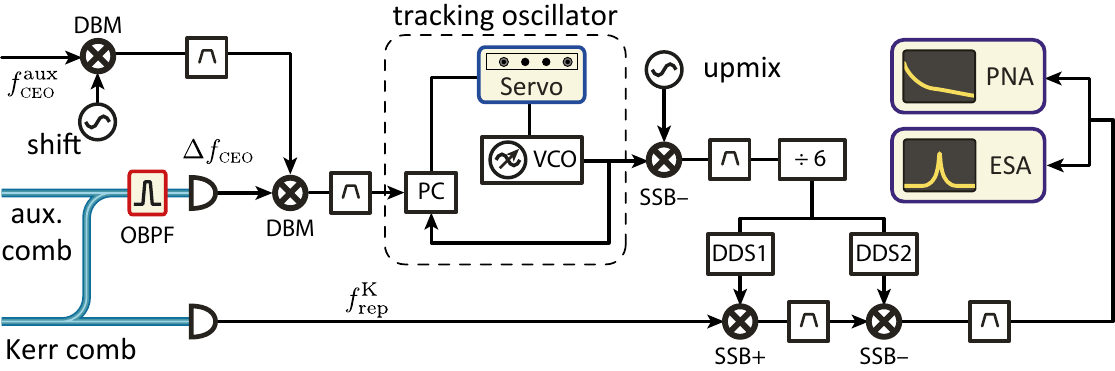}
	\caption{\textbf{Frequency chain for the transfer oscillator division}
		Implementation of the optical-to-microwave frequency division using the 2-DDS transfer oscillator scheme. The $\frep$ component of the Kerr comb detected with a fast photodiode (lower path) is mixed with the CEO signal frequency-divided by a factor $N$ (upper path). The division from 15 GHz to 1.095 MHz is realised by a frequency pre-scaler ($\div6$) followed by two DDS in parallel that output signals at 100 MHz and 101.095 MHz, respectively, from the 2.5 GHz input clock signal. The CEO frequency of the Kerr comb is indirectly obtained from the subtraction of the frequency-shifted auxiliary self-referenced comb CEO $\fauxceo$ with the optical beat-note from the two combs ($\Delta \fceo$), due to the fact that the phase noise of the repetition rate of the two combs is  correlated by an injection locking scheme. DBM, double-balanced mixer; VCO, voltage-controlled oscillator; PC, digital phase comparator; SSB(+/-), single sideband mixer (sum/difference frequency); DDS, direct digital synthesiser; PNA, phase noise analyser; ESA, electrical spectrum analyser.
	}
	\label{fig:DivisionChain}
\end{figure*}

\section{Division chain}
\label{SI:TO_chain}
The transfer oscillator approach is implemented in this work with a 2-DDS scheme as introduced by Brochard and co-authors~\cite{Brochard2018} to perform electronic division with a finely adjustable ratio.  This implementation also makes possible the generation of a low-noise single-tone RF output signal by efficiently filtering out other spurious peaks that would occur with a single DDS. The practical realisation of this scheme with the Kerr comb is depicted in \fref{fig:DivisionChain}.

The beat signal $ \Delta \fceo $ between the two combs is mixed in a double balanced mixer (DBM) with the CEO signal of the auxiliary comb $ \fauxceo $ (detected using a standard $f-2f$ interferometer) in order to remove this contribution and retrieve $ \fkceo $. Prior to this mixing, the CEO signal of the auxiliary comb, which is stabilised at 20~MHz, is frequency-up-shifted using a low-noise synthesiser. This provides a straightforward means to arbitrarily tune $\fauxceo$ without changing it optically (which would also change the optical beat-note frequency by $\Delta \fceo$) and without frequency noise degradation, as the measured phase noise of the intermediate synthesiser is comparatively negligible. Thereby, the effective sign of $\fauxceo$ can be quickly changed, without changing any RF component, in order to properly remove its contribution when mixing with $\Delta \fceo$ in the DBM.

\begin{figure}[h!bt]
	\centering
	\includegraphics[width=\columnwidth]{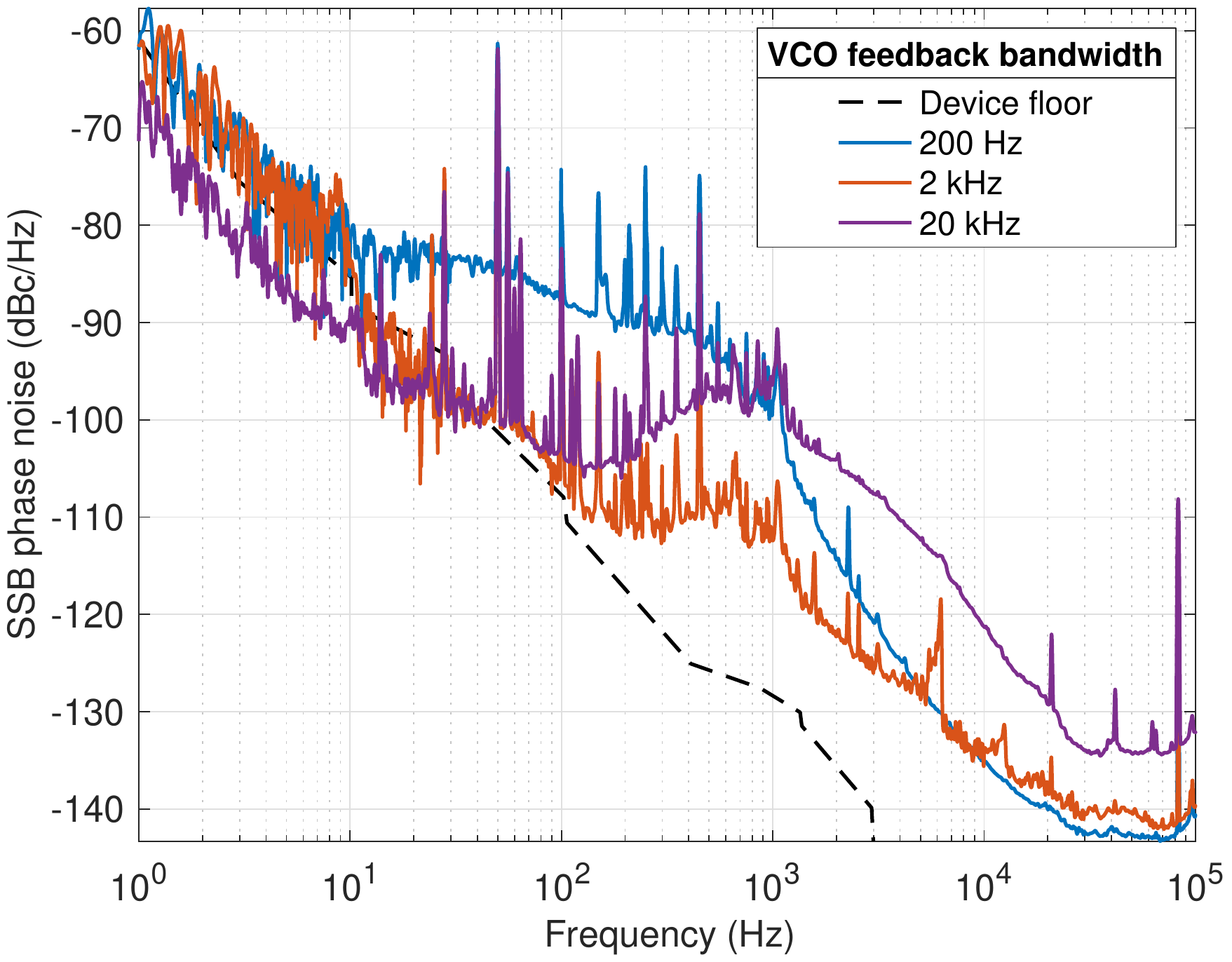}
	\caption{\textbf{Tracking oscillator effect.}
		Influence of the feedback bandwidth of the tracking oscillator, used to filter the Kerr comb $\fceo$, onto the final generated RF signal. The lowest phase noise of the final RF signal is obtained for a locking bandwidth of the VCO of $\sim 2$~kHz (red curve), similar to the bandwidth of the injection locking process used to correlate the noise of the repetition rate of the two combs. A lower ($\sim 200$~Hz, blue curve) or higher ($\sim 20$~kHz, purple curve) bandwidth results in a higher noise of the generated RF signal.
	}
	\label{fig:VCOEFFECT}
\end{figure}

Furthermore, this additional flexibility enables us to finely adjust the signal frequency at the DBM output, which contains the effective Kerr comb CEO, to make it coincide with the frequency of a tracking oscillator around 40~MHz. This tracking oscillator consists in a narrow-band low-noise voltage-controlled oscillator (VCO) that is phase-locked to $\fkceo$. The fine adjustment of the VCO locking bandwidth (by tuning the PI parameters of the feedback) enables us to filter the noise of the detected $\fkceo$ in order to match the $\sim 2$ kHz injection locking bandwidth over which the relation $\fkrep = 56 \fauxrep$ is ensured. Thereby, the contribution of the residual uncorrelated fluctuations between the two combs in the generated ultralow-noise RF signal is minimised. Lower ($\sim 200$~Hz) or higher ($\sim 20$~kHz) feedback bandwidths lead to an increased noise in the final signal at low or high Fourier frequencies respectively (\fref{fig:VCOEFFECT}) as a result of the imperfect  compensation of the auxiliary comb noise. Furthermore, as an RF signal with a sufficient signal-to-noise ratio (SNR) of more than 30~dB is needed for a proper and stable operation of the subsequent frequency divider, this tracking oscillator helps improving the signal quality, so that even a fairly low SNR of the signal at the output of the DBM makes possible to implement the transfer oscillator method.

\begin{figure}[ht]
	\centering
	\includegraphics[width=\columnwidth]{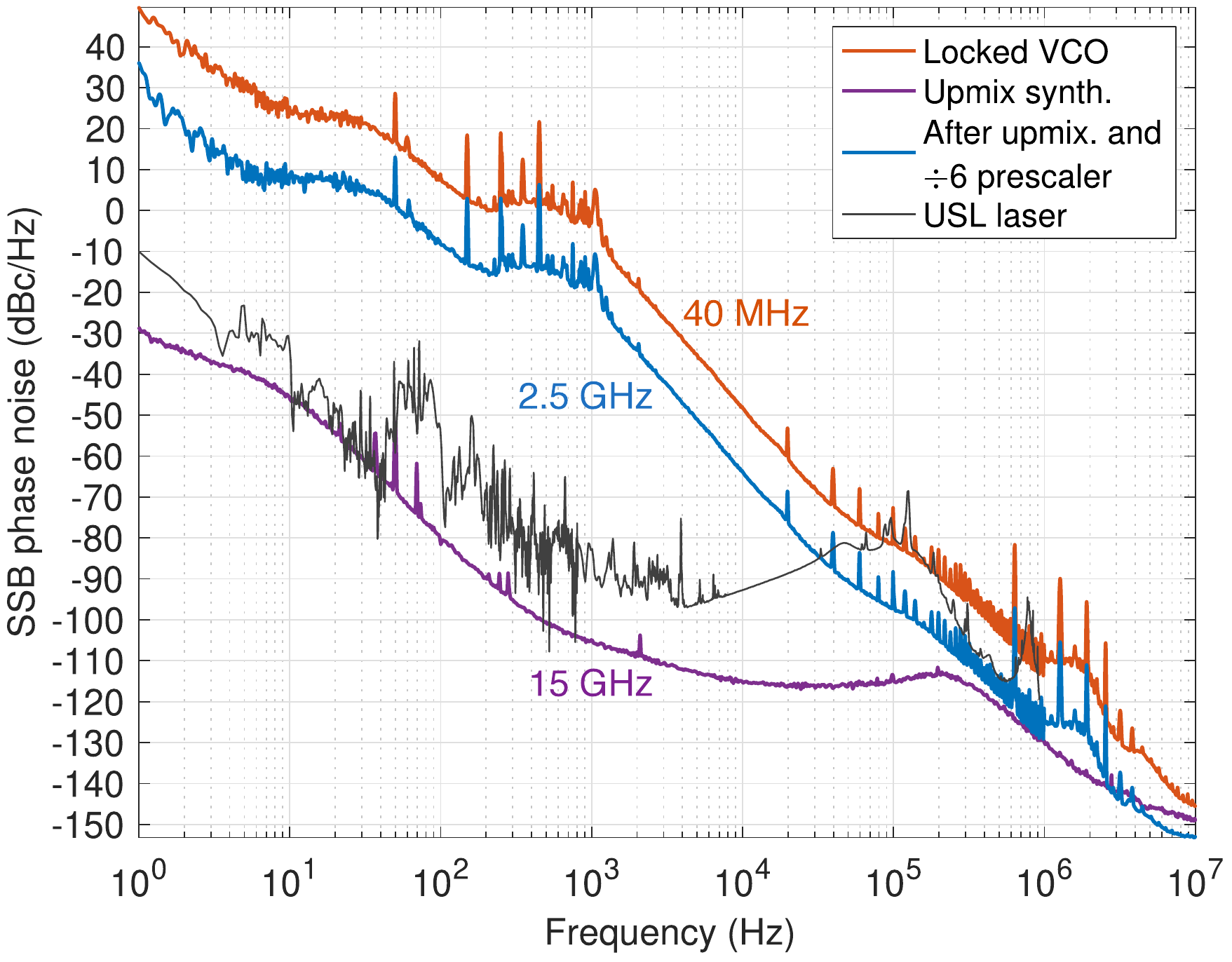}
	\caption{\textbf{Upmixing and pre-scaler division.}
		Evolution of the signal phase noise during the upmixing and pre-scaler division of the filtered Kerr comb CEO. The carrier frequencies are indicated on the plot. The noise of the upmixing synthesiser (purple curve) is lower than the USL noise (black), such that it does not limit the final division result. Upmixing with the comb repetition rate would avoid this potential limitation.
	}
	\label{fig:UPMIX}
\end{figure}

The signal after the tracking oscillator is up-converted to 15~GHz using a synthesiser with an absolute phase noise lower than the USL (see \fref{fig:UPMIX}), so that its noise has a negligible contribution in the final signal. This frequency shift is necessary to perform the subsequent frequency division by a large number of around 13,698. Eventually, this synthesiser could be replaced by the repetition rate of the Kerr-comb to alleviate any of the associated limitation. This was not implemented here due to the lack of appropriate filters.

The frequency division from 15~GHz to 1.095~MHz is realised first by a frequency pre-scaler ($\div$6, RF Bay FPS-6-15) followed by two DDS (AD9915 evaluation board) in parallel that respectively output signals at 100~MHz and 101.095~MHz from their 2.5~GHz input clock signal. These two signals are subsequently mixed with the repetition rate of the Kerr comb separately detected using a fast photodiode (Discovery Semiconductors DSC40) and filtered to select the proper component that corresponds to $f_{\rm RF} = \fkrep + \fkceo/N = \nu_{\textsc{usl}}/N$. The sequential mixing followed by filtering after each DDS allows for the efficient rejection of the spurious peaks occurring at harmonics of the DDS signals, thanks to their relatively high frequency spacing ($100$~MHz range). 

\begin{figure}[ht]
\centering
\includegraphics[width=\columnwidth]{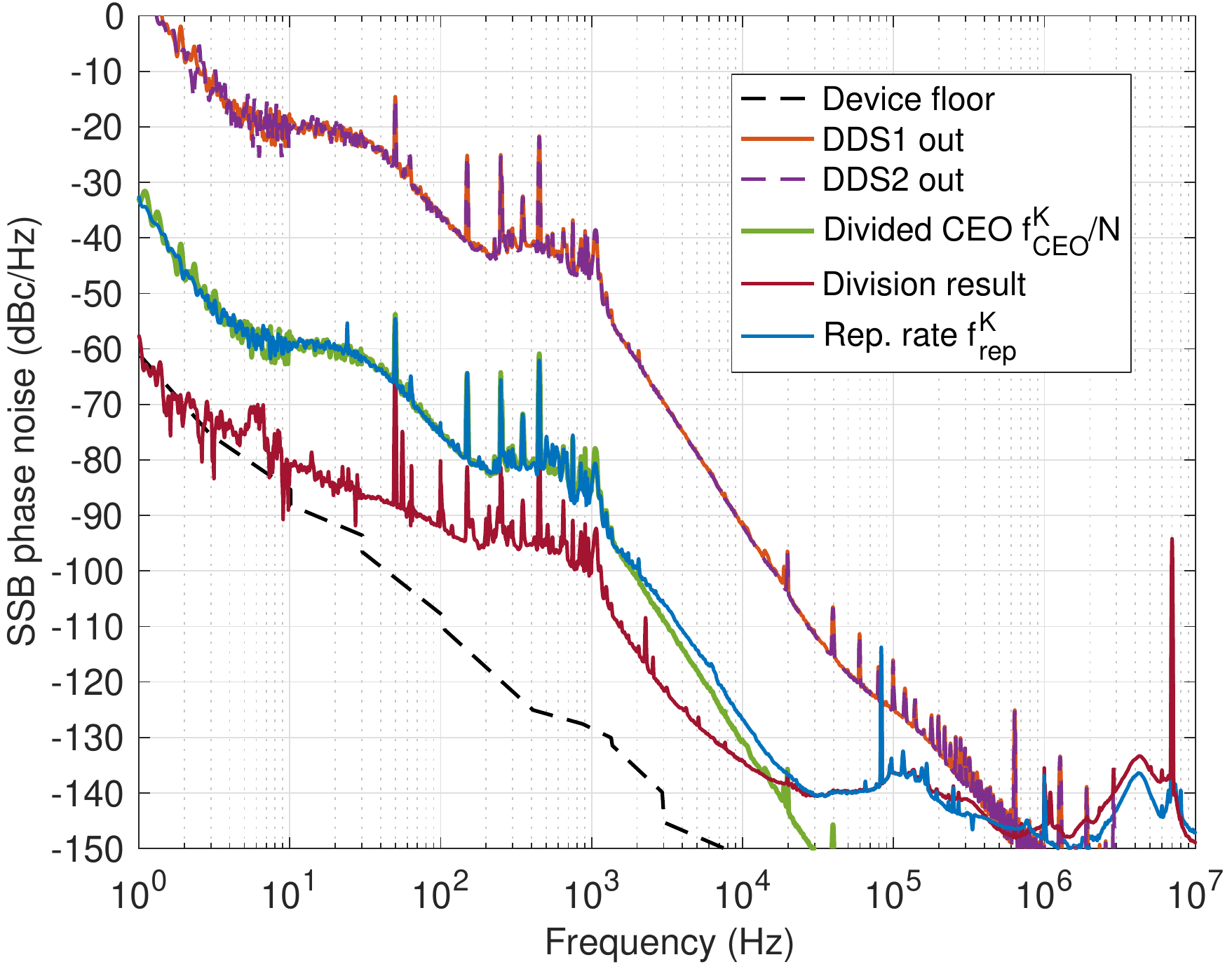}
\caption{\textbf{Noise division demonstration}
Demonstration of the noise division achieved by the 2-DDS scheme. The orange and dashed violet curves show the phase noise PSD separately measured at the output of each DDS at 101.095 MHz (DDS1) and 100 MHz (DDS2), respectively. The green curve displays the noise of the frequency-divided CEO signal of the Kerr comb, which overlaps the noise of the repetition rate (blue curve). Therefore, these two noise contributions compensate each other to a large extend in the final RF signal (red curve), which demonstrates the noise improvement brought by the transfer oscillator scheme limited here by the Kerr comb injection locking bandwidth.
}
\label{fig:DIVISION}
\end{figure}

The generated ultralow noise RF signal is characterised using a phase noise analyser (PNA, model FSWP26 from Rohde-Schwarz) and an electrical spectrum analyser (ESA, model FSW43 from Rohde-Schwarz).
The effect of the frequency division performed with the 2-DDS scheme is illustrated in \fref{fig:DIVISION}. The frequency difference of the two signals at 100~MHz and 101.095~MHz, respectively, gives a signal at 1.095~MHz which correspond to $\fkceo/N$ and is strongly correlated with $\fkrep$ at 14.09~GHz. Mixing these two signal results in the generation of a very low noise RF signal that demonstrates a high rejection of the Kerr comb phase noise.

\section{Sign effect in the transfer oscillator}
\label{SI:SignEffect}
\begin{figure}[h]
\centering
\includegraphics[width=\columnwidth]{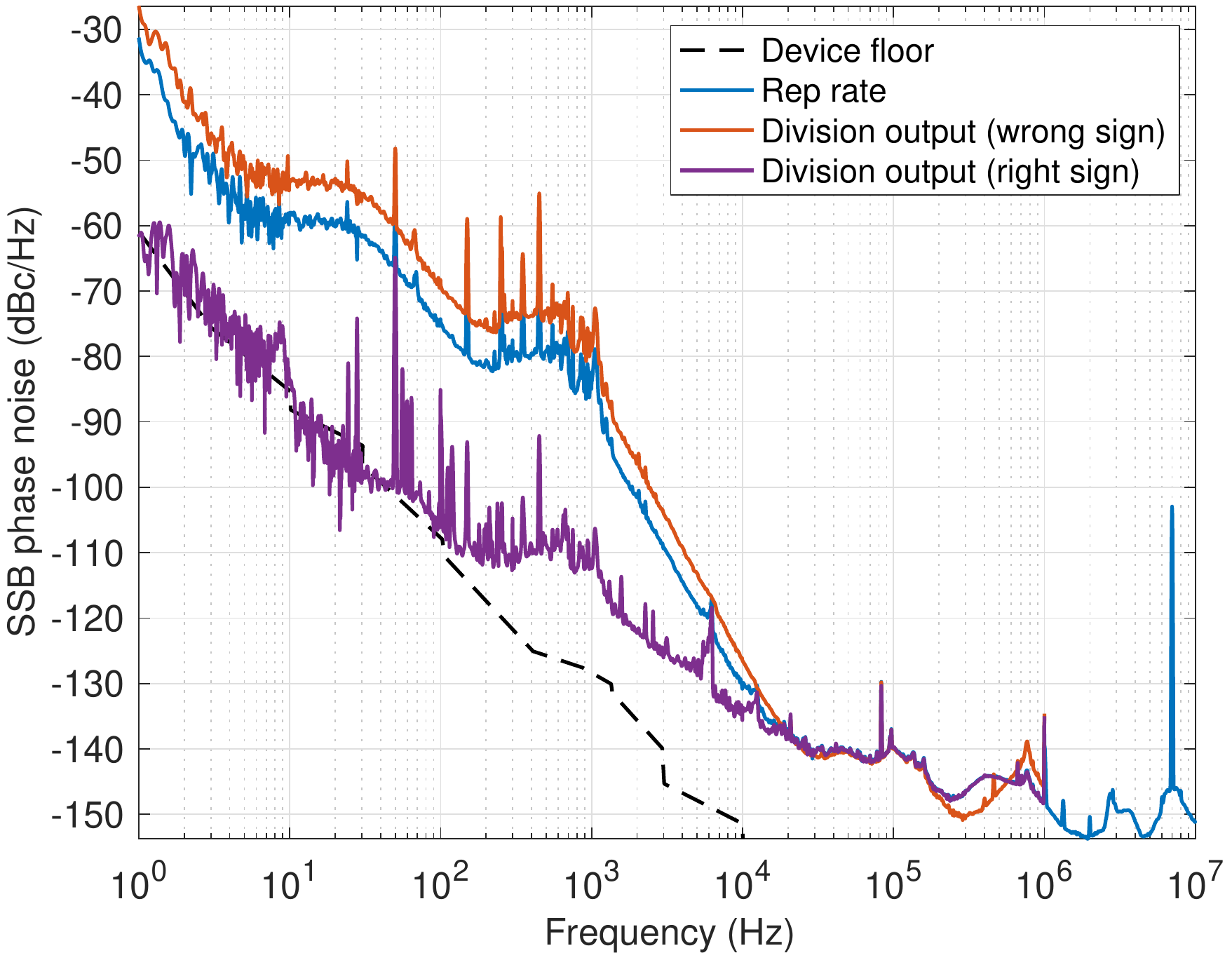}
\caption{\textbf{Sign effect}
Demonstration of the adjustment of the sign of the correction of the Kerr comb $\fkrep$ noise in the low-noise RF output signal generated by the transfer oscillator scheme. The blue curve displays the phase noise of the Kerr comb repetition rate. The purple curve shows the phase noise of the generated RF signal obtained with the correct sign where the $\fkrep$ noise is removed, whereas the orange curve corresponds to the other sign (obtained by inverting the frequency of the two DDS), which leads to a 6 dB noise increase (the output signal contains twice the frequency fluctuations of $\fkrep$).
}
\label{fig:SIGNEFFECT}
\end{figure}

The high rejection of the Kerr comb phase noise offered by the transfer oscillator scheme requires mixing signals with the proper sign combination, so that the frequency fluctuations of $\fkrep$ and $\fkceo$ are indeed compensated in the final RF signal. The sign of $\fkceo$ is determined by the heterodyne beat between the Kerr comb and the auxiliary comb, which can be controlled by the repetition rate of the auxiliary comb, and by the subtracted CEO signal of the auxiliary comb, whose sign can be changed using the frequency-shifting synthesiser as previously mentioned. The sign of the $\fkrep$ contribution to be removed can be adjusted by inverting the output frequencies of the two DDS, without changing any RF component. This is illustrated in \fref{fig:SIGNEFFECT}, which shows how the noise is correctly compensated with the proper sign and increases by a factor of 4 (in terms of PSD, or +6 dB) compared to the noise of $\frep$ with the incorrect sign (as the resulting signal corresponds to $\nu_{\textsc{usl}}/N + 2 \frep$).

\section{Source of imperfect noise compensation}
\label{SI:DelayCompensationEffect}
The noise compensation in the transfer oscillator method relies on the subtraction of various noise contributions of the micro-resonator comb. Perfect noise compensation occurs when no relative delay is introduced between $\fceo/N$ and $\frep$ at the time of their final mixing. If a significant delay occurs between the signals, the noise compensation may be incomplete as the signals become imperfectly correlated~\cite{Rubiola2005a} and the residual (uncompensated) noise scales according to 
\begin{equation}
S_\varphi^{\rm signal}(f) =  \dfrac{1}{N^2} \, S_\varphi^{\textsc{usl}}(f) +  4 \sin^2\left( \pi \tau f \right) \, S_\varphi^{\rm rep}(f) \, ,
\end{equation}
where $\tau$ is the relative delay, $f$ is the Fourier frequency and $S_\varphi$ denotes the phase noise power spectral density. If $\tau = 0$ the phase noise of the repetition rate is properly cancelled. Some care is thus needed to minimise the delays, in order to maximise the noise cancellation bandwidth, but this factor is not critical. For example, a coarse length mismatch of 10~m would correspond to a delay
of $\sim 42$~ns (assuming a velocity factor of 80~\%), resulting in a fully uncompensated noise (0~dB rejection) reached at a Fourier frequency of $\sim4$~MHz.

\bigbreak
\def\bibsection{}  
\noindent\textbf{Supplementary References }
\medbreak
\bibliography{library}